\documentclass[aps,prb,twocolumn,superscriptaddress,floatfix,longbibliography]{revtex4-2}
\usepackage[english]{babel}
\usepackage[utf8]{inputenc}
\usepackage{mathtools}
\usepackage{physics}
\usepackage{float}
\usepackage{braket}
\usepackage{bbm}
\usepackage{bm}
\usepackage{amsbsy}
\usepackage{amsthm}
\usepackage{amssymb}
\usepackage{amsfonts}
\usepackage{amsmath}
\usepackage{soul}
\setstcolor{red}
\usepackage{dsfont} 
\usepackage{graphicx} 
\usepackage{hhline}
\usepackage{epsfig}
\usepackage{epstopdf}
\usepackage{dsfont}
\usepackage{xcolor}
\usepackage{color}
\usepackage{verbatim}
\usepackage{nomencl}
\usepackage[colorlinks]{hyperref}
\usepackage{float}
\usepackage[normalem]{ulem}
\usepackage{multirow}
\usepackage{makecell}
\usepackage{url}

\DeclareUnicodeCharacter{0301}{\'{e}}

\begin{document}

\title{Dynamical onset of quasiprobability negativity in quantum many-body systems}
\author{Rohit Kumar Shukla}
\email[]{rohitkrshukla.rs.phy17@itbhu.ac.in}
\affiliation{Department of Chemistry, Bar-Ilan University, Ramat-Gan 52900, Israel}
\affiliation{Institute of Nanotechnology and Advanced Materials, Bar-Ilan University, Ramat-Gan 52900, Israel}
\affiliation{Center for Quantum Entanglement Science and Technology, Bar-Ilan University, Ramat-Gan 52900, Israel}

\author{Amikam Levy}
\email[]{amikam.levy@biu.ac.il }
\affiliation{Department of Chemistry, Bar-Ilan University, Ramat-Gan 52900, Israel}
\affiliation{Institute of Nanotechnology and Advanced Materials, Bar-Ilan University, Ramat-Gan 52900, Israel}
\affiliation{Center for Quantum Entanglement Science and Technology, Bar-Ilan University, Ramat-Gan 52900, Israel}

\date{\today}

\begin{abstract}
Time-dependent quasiprobability distributions provide a quasiprobabilistic
description of sequential measurement statistics generated by quantum dynamics, and can reveal nonclassical features with no classical probabilistic
counterpart. Yet the dynamical emergence of their negativity in many-body systems remains largely unexplored. We introduce the \emph{first-time negativity} (FTN) of the Margenau-Hill quasiprobability as a dynamical indicator of when local measurement sequences in an interacting quantum system begin to exhibit genuinely nonclassical behavior.
Using the Ising chain, we show that FTN discriminates clearly between interaction-dominated and field-dominated regimes, is systematically reshaped by temperature, and responds sensitively to the breaking of integrability. For spatially separated measurements, FTN appears abruptly near the
interaction-field crossover, while measurements at opposite boundaries
exhibit a distinct weak-field branch whose onset time grows with chain length
and is consistent with ballistic propagation across the system.
We further compare the numerical onset of negativity with a recently proposed quantum speed limit (QSL) for quasiprobabilities, which provides a geometric benchmark for the observed dynamics. 
Our results identify FTN as a practical and experimentally accessible probe of the real-time onset of quasiprobability negativity and contextual sequential measurement statistics, directly suited to current platforms capable of sequential weak and strong measurements.
\end{abstract}

\maketitle

\section{Introduction}

 Quasiprobability distributions extend classical probability theory to describe
genuinely quantum features such as interference, contextuality, and the effects
of noncommuting observables~\cite{wigner1932quantum,kirkwood1933quantum,
dirac1945analogy,lostaglio2023kirkwood,arvidsson2024properties,
pusey2014anomalous,kunjwal2019anomalous}. Unlike ordinary probabilities,
quasiprobability representations may take negative values, and in some cases
complex values, with the precise operational properties depending on the
representation.

In continuous-variable systems, this role is played by phase-space
distributions such as the Wigner, Glauber-Sudarshan, and Husimi
functions~\cite{wigner1932quantum,glauber1963coherent,
sudarshan1963equivalence,husimi1940some}, whose negativity has long been
recognized as a marker of nonclassicality~\cite{walls1994atomic}. For discrete
systems, the Kirkwood-Dirac quasiprobability (KDQ) provides an analogous joint
description for generally incompatible observables~\cite{kirkwood1933quantum,
dirac1945analogy,lostaglio2023kirkwood,arvidsson2024properties}. Its real
part, the Margenau-Hill quasiprobability (MHQ)~\cite{margenau1961correlation},
retains the correct single-observable marginals for the two sequentially
measured observables considered below. The KDQ and MHQ have been developed and
employed in foundational, thermodynamic, and information-theoretic
settings~\cite{barut1957distribution,johansen2007quantum,
allahverdyan2014nonequilibrium,lostaglio2018quantum,
yunger2018quasiprobability,gonzalez2019out,kunjwal2019anomalous,
levy2020quasiprobability,arvidsson2021conditions,pei2023exploring,
stepanyan2024energy,wagner2024quantum}, and can be reconstructed
experimentally using weak-then-strong measurement
schemes~\cite{monroe2021weak,lostaglio2023kirkwood,wagner2024quantum}.

For KD and MH quasiprobabilities, nonpositivity provides a particularly sharp
signature of nonclassicality. Noncommutativity alone does not guarantee
negativity~\cite{arvidsson2021conditions,de2021complete}; in the weak-measurement
protocols that operationally realize the KD distribution, observing anomalous or
negative KD/MH values are tightly linked to contextuality, in the sense that the resulting sequential statistics cannot be reproduced by any noncontextual
hidden-variable model under standard noninvasiveness
assumptions~\cite{pusey2014anomalous,kunjwal2019anomalous}.
The imaginary part of the KDQ captures measurement back-action and disturbance~\cite{jozsa2007complex,hofmann2011role,dressel2012significance,monroe2021weak,budiyono2023quantifying,hernandez2024interferometry}.
Moreover, the KD/MH negativity has been connected to enhanced performance in metrology and thermodynamics~\cite{lostaglio2020certifying,jenne2021quantum,allahverdyan2014nonequilibrium,levy2020quasiprobability,mallik2025probing}, to work statistics in many-body systems~\cite{santini2023work,francica2023quasiprobability,gherardini2024quasiprobabilities}, and to nonclassical signatures in projective-measurement protocols~\cite{hernandez2024projective,yunger2018quasiprobability,gherardini2024quasiprobabilities}.

However, recent work shows that KD negativity is not, in general, a universal
criterion for contextuality in arbitrary representations, so its status as a
contextuality witness must always be understood in relation to a specific
operational scenario~\cite{arvidsson2024properties,schmid2024kirkwood}.
In this sense, quasiprobabilities provide a compact language for quantifying
“how quantum’’ a given dynamical process is, beyond what can be inferred from
standard correlation functions alone.

A natural yet largely unexplored question is \emph{when} such nonclassicality
first appears in time. In other words, given a many-body system, a choice of local observables, and an initial state, what is the earliest time at which the corresponding quasiprobability distribution must cease to be nonnegative? 
 Although simple and natural, this question has not, to the best of our
knowledge, been formulated as a dynamical diagnostic in interacting many-body
systems. This motivates the first-time negativity introduced here, which turns
the onset of negativity into an operational timescale and provides a natural
target for controlling when nonclassicality emerges.
Recent work has derived a QSL that bounds from below the time at which KDQ entries can become nonpositive~\cite{pratapsi2025quantum}. However, this bound is constructed for a single pair of projectors and need not be saturated by the actual dynamics; in particular, it may be finite even when the associated quasiprobability remains nonnegative for all times. This suggests that the true onset of negativity is controlled not only by geometric constraints, but also by the detailed structure of the Hamiltonian and the chosen observables.

In this work, we address this dynamical question in a paradigmatic interacting system: a one-dimensional Ising chain with transverse and longitudinal fields. We focus on local, experimentally accessible probes and on the Margenau-Hill quasiprobability associated with sequential measurements of single-qubit Pauli operators. As our main diagnostic we introduce the \emph{first-time negativity} (FTN), $t_{\rm FTN}$, defined as the earliest time at which any MH entry becomes negative for a given pair of local projectors. This quantity provides a sharp operational timescale for the onset of contextual quantum interference in sequential measurements. We study how $t_{\mathrm{FTN}}$ depends on the transverse field, on temperature, on integrability breaking by a longitudinal field, and on the spatial separation between measurement sites.

Our analysis reveals several robust features. In the integrable transverse-field Ising model, initialized in its ground state, MH negativity appears only for $\sigma_z$ probes. The corresponding on-site FTN exhibits distinct asymptotic scalings in the interaction-dominated and field-dominated regimes and depends only weakly on the total system size over the range studied, confirming the local character
of the single-site onset. Finite temperature progressively broadens the sharp zero-temperature feature around the quantum critical point into a finite-temperature crossover and eventually suppresses negativity in the high-temperature, fully mixed limit. Comparison with the KDQ-based QSL shows that the bound captures only the maximal kinematic rate of change: it can predict a finite minimal time to \emph{possible} nonpositivity even in regimes where the MH quasiprobability remains nonnegative at all times, whereas the FTN directly tracks the actual emergence of negativity.

Extending the construction to projectors at different lattice sites reveals
two distinct spatial behaviors. For generic separated sites, no negative MH
entry is observed in the weak-field regime within the accessible time window,
and a finite FTN appears abruptly near the interaction--field crossover.
Measurements at opposite open boundaries exhibit a distinct weak-field branch
whose onset time grows approximately linearly with chain length and inversely
with the transverse-field strength, consistent with a ballistic
boundary-to-boundary propagation timescale. In the field-dominated regime,
the onset at distant sites becomes less sensitive to the transverse field and
retains a pronounced dependence on separation.

Breaking integrability by adding a longitudinal field lifts the
$\mathbb{Z}_2$ spin-flip symmetry that protects the $\sigma_x$ channel in the
integrable model, so that both $\sigma_z$ and $\sigma_x$ observables develop
negative quasiprobability entries.

These results show that the FTN of the Margenau-Hill quasiprobability provides a concise and experimentally relevant measure of when
and where quasiprobability negativity first emerges in many-body dynamics. In the weak-measurement scenario, and under the standard noninvasiveness
assumptions described above, it also identifies when the resulting sequential statistics become incompatible with a noncontextual hidden-variable
model~\cite{pusey2014anomalous,kunjwal2019anomalous}. FTN therefore complements more traditional diagnostics such as correlators and out-of-time-ordered
functions. More broadly, it provides a natural control objective, opening the possibility of advancing, delaying, or suppressing the emergence of negativity
through Hamiltonian engineering, state preparation, or measurement design.

The remainder of this paper is organized as follows.  
In Sec.~\ref{sec:model} we introduce the Ising spin model and define the
Margenau-Hill quasiprobability, its negativity, and the associated first-time
negativity $t_{\mathrm{FTN}}$, together with the relevant quantum speed-limit
bound.  
Section~\ref{Result} presents our main results. We first analyze the integrable transverse-field case at zero and finite temperature, considering both local measurements and measurements on spatially separated sites. We then study the nonintegrable regime in the presence of a longitudinal field.  
In Sec.~\ref{Conclusion}, we summarize our findings and discuss possible
extensions to other models.

\section{Theoretical framework}
\label{sec:model}

\subsection*{Ising spin system}

We consider a one-dimensional Ising chain of $N$ qubits with open boundaries, described by
\begin{equation}
\label{eq:H}
 H = -J \sum_{n=1}^{N-1} \sigma_n^x \sigma_{n+1}^x
        - h_z \sum_{n=1}^{N} \sigma_n^z
        - h_x \sum_{n=1}^{N} \sigma_n^x,
\end{equation}
where $J$ denotes the nearest-neighbour exchange coupling and $h_z$ and $h_x$
are the transverse and longitudinal fields, respectively. 
The sign of $J$
distinguishes the ferromagnetic ($J>0$) and antiferromagnetic ($J<0$) models,
which can lead to quantitative (and in some regimes qualitative) differences in
the dynamics. In this work, we focus on the ferromagnetic case; the implications
of $J<0$ are discussed where relevant in the main text and treated in more detail in the Appendix. We set $\hbar=1$ throughout.

This model captures the competition between ferromagnetic exchange along the $x$ direction, a transverse field polarizing the spins along $z$, and an additional longitudinal field aligning them along $x$. 
The interplay between the transverse field $h_z$ and the exchange coupling $J$ governs the magnetic phase structure: for $h_x = 0$, the model reduces to the integrable transverse-field Ising model (TFIM), which exhibits a quantum phase transition at $h_z/J = 1$, separating a ferromagnetic ground state ($h_z/J < 1$) from a paramagnetic one ($h_z/J > 1$)~\cite{sachdev1999quantum}.
Introducing a finite longitudinal field $h_x$ breaks the integrability of the transverse-field Ising model, rounding the sharp transition into a crossover \cite{noh2021operator}.  The resulting dynamics depend sensitively on the full set of parameters $(J, h_z, h_x)$, and are generically chaotic away from fine-tuned limits or low-energy sectors. 

In the following, we will often focus on local qubit operators such as $\sigma_i^z$. Their two-time correlations and associated quasiprobabilities provide a direct probe of the distinct dynamical regimes of the model.

\subsection*{Quasiprobability distributions and negativity}

Local, single-site probes are directly accessible in today’s quantum platforms, 
including trapped ions~\cite{bruzewicz2019trapped}, Rydberg atom arrays~\cite{browaeys2020many}, 
superconducting circuits~\cite{elder2020high}, and NV centers in diamond~\cite{cambria2025scalable}. 
Motivated by this, we diagnose dynamical nonclassicality using a \emph{local} two-time 
quasiprobability that can be reconstructed from weak-then-strong sequential measurements 
on a single qubit.

Let $V$ and $W$ be observables with projectors $\{\Pi_\gamma\}$ and $\{\Xi_\delta\}$ on a single qubit. When acting on sites $m$ and $n$ of an $N$-qubit chain, the projectors are embedded into the full Hilbert space as
$\Pi^m_\gamma \equiv \mathbb{I}^{\otimes m-1} \!\otimes \Pi_\gamma \otimes \mathbb{I}^{\otimes N-m}$ and
$\Xi^n_\delta \equiv \mathbb{I}^{\otimes n-1} \!\otimes \Xi_\delta \otimes \mathbb{I}^{\otimes N-n}$.
Following the standard KD construction for ordered pairs of projectors
\cite{lostaglio2023kirkwood,
arvidsson2024properties}, we define the two-time KD quasiprobability for the
ordered sequence ``measure $W$ at $t{=}0$'' and then ``measure $V$ at time
$t$'' as
\begin{equation}
p^{mn}_{\gamma\delta}(t) \;=\; \mathrm{Tr}\!\big[\,\Pi^m_\gamma(t)\,\Xi^n_\delta\,\rho_0\,\big], 
\quad \Pi^m_\gamma(t)=e^{iHt}\Pi^m_\gamma e^{-iHt}.
\label{eq:KD}
\end{equation}
Here, $\rho_0$ is the initial state of the many-body system, taken as the ground state of the Hamiltonian. 
The real part of the KD quasiprobability, 
\begin{equation}
 q^{mn}_{\gamma\delta}(t)\equiv \Re\,p^{mn}_{\gamma\delta}(t),   
\end{equation}
is the MHQ~\cite{margenau1961correlation}, which will be our focus. 
Although not a genuine probability distribution, $q^{mn}_{\gamma\delta}(t)$ has consistent marginals: summing over one index yields the correct distribution for the other observable, and weighted sums reconstruct expectation values and two-time correlators. 
When the two projectors commute, $q^{mn}_{\gamma\delta}(t)\ge 0$ and coincides with an ordinary joint probability. 
By contrast, \emph{negative} values can arise only if the sequential measurements are incompatible; such negativity reflects interference between different time orderings and, under standard operational assumptions (in particular, that weak pre-measurements are operationally noninvasive)~\cite{pusey2014anomalous,kunjwal2019anomalous}, rules out any noncontextual hidden-variable model reproducing the same sequential statistics.

For an $N$-qubit chain of Hilbert-space dimension $\mathcal{D}=2^N$, each local projector 
$\Pi_\gamma^m$ or $\Xi_\delta^n$ has rank $\mathcal{D}/2$, since it acts only on a single site. 
Consequently, the KD elements $q^{mn}_{\gamma\delta}(t)$ 
capture interference between $\mathcal{D}/2$-dimensional subspaces rather than rank-1 rays. 
This coarse-grained setting is natural for local qubit readout and still allows quasiprobability 
negativity, although coarse-graining can reduce the observed magnitude of negativity compared with a fully nondegenerate (rank-1) refinement.

We quantify the extent of negativity~\cite{gonzalez2019out,arvidsson2021conditions} 
\begin{equation}
\mathcal{N}^{mn}(t)\;=\;\sum_{\gamma,\delta}\!\Big|\,q^{mn}_{\gamma\delta}(t)\Big| \;-\; 1,
\label{eq:neg}
\end{equation}
and define the \emph{first–time negativity} (FTN) as the earliest time when any MH entry turns negative,
\begin{equation}
t_{\mathrm{FTN}}^{mn}\;=\;\min\!\Big\{t>0:\ \exists\,\gamma,\delta\ \text{with}\  q^{mn}_{\gamma\delta}(t)<0\Big\}.
\label{eq:ftn}
\end{equation}

\subsubsection*{First-time negativity and the quantum speed limit.}
The central dynamical scale we extract is the first-time negativity
$t_{\mathrm{FTN}}^{\,mn}$ in Eq.~\eqref{eq:ftn}, i.e., the earliest instant when a quasiprobability entry becomes negative. This quantity is useful because
it turns quasiprobability negativity from a static yes or no property into a dynamical timescale. In this sense, FTN plays a role analogous to other onset
times in many-body dynamics, such as relaxation, decoherence, or scrambling times: it asks how long the system can evolve before the sequential measurement
statistics cease to admit a nonnegative quasiprobabilistic description.

For a fixed initial state, pair of local observables, and measurement ordering, $t_{\mathrm{FTN}}^{mn}$ provides an operational timescale for comparing how rapidly different Hamiltonian regimes, temperatures, or spatial separations generate nonclassical statistics. It also has a direct experimental meaning: $t_{\mathrm{FTN}}^{mn}$ sets the minimum time window over which coherent dynamics and sequential measurements must be resolved in order to observe quasiprobability negativity.
Recent work~\cite{pratapsi2025quantum} has derived a QSL for nonpositivity of KD quasiprobabilities, which provides a lower bound on $t_{\mathrm{FTN}}^{mn}$.  Since the local projectors are degenerate, we first normalize the
time-evolved projector to unit trace. Defining
$
r_\gamma
=
\operatorname{Tr}\!\left[\Pi_\gamma^m\right]
=
\frac{\mathcal{D}}{2},
$ and $ 
\widetilde{\Pi}_\gamma^m(t)
=
\frac{\Pi_\gamma^m(t)}{r_\gamma},
$
the corresponding
normalized MH entry is
\begin{equation}
\widetilde q_{\gamma\delta}^{mn}(t)
=
\operatorname{Tr}\!\left[
\rho_\delta^n\widetilde{\Pi}_\gamma^m(t)
\right]
=
\frac{q_{\gamma\delta}^{mn}(t)}{r_\gamma},
\qquad
\rho_\delta^n
=
\frac{1}{2}
\left\{
\rho_0,\Xi_\delta^n
\right\}.
\end{equation}
Because $r_\gamma>0$, the normalized and unnormalized quasiprobabilities
cross zero at the same time.

In our notation, the QSL for possible nonpositivity reads
\begin{equation}
T_{\gamma\delta;mn}^{\mathrm{re}}
=
\frac{
\tau\!\left(\rho_\delta^n,0\right)
-
\tau\!\left(
\rho_\delta^n,
\widetilde q_{\gamma\delta}^{mn}(0)
\right)
}{
\Delta L_\gamma^m
},
\label{eq:Tre}
\end{equation}
where $\Delta L_\gamma^m$ is the standard deviation of the symmetric
logarithmic derivative associated with the normalized projector
$\widetilde{\Pi}_\gamma^m(t)$. The function
$\tau(\rho_\delta^n,x)$ is the arccosine parametrization introduced in
Ref.~\cite{pratapsi2025quantum}; it maps an allowed expectation value $x$ to
an angular coordinate within the interval defined by the minimum and maximum
eigenvalues of the Hermitian operator $\rho_\delta^n$. Its explicit
definition and the numerical evaluation of the bound are given in
App.~\ref{App:QSL}.

The quantity $T_{\gamma\delta;mn}^{\mathrm{re}}$ is a lower bound on the
time at which the corresponding MH entry can reach zero. A finite value does
not imply that the actual quasiprobability becomes negative. We therefore use
the QSL only as a geometric benchmark, whereas
$t_{\mathrm{FTN}}^{mn}$ identifies the negativity actually generated by the
Ising-chain dynamics.

\subsubsection*{Quantumness beyond correlations}

Two-time correlators such as $C_{VV}(t)=\langle V(t)V(0)\rangle$ diagnose
memory and relaxation and can often be reproduced by classical stochastic
dynamics with suitable kernels~\cite{carballeira2021stochastic}. By contrast, MH negativity does not quantify the magnitude of correlations; rather, in the usual weak-measurement implementation it certifies that the two sequential
outcomes cannot be described by any single, nonnegative, context-independent
joint distribution~\cite{pusey2014anomalous,schmid2024kirkwood}. Consequently,
$\mathcal{N}^{mn}(t)$ and $t_{\mathrm{FTN}}^{mn}$ track the onset of
contextual quantum interference in the measurement statistics, whereas the
correlator $C_{VV}(t)$ reflects the persistence of classical memory.
In the Ising chain, these scales need not coincide: symmetry, locality, and
operator spreading can delay the onset of negativity even when $C_{VV}(t)$ has
already decayed (or, in some regimes, allow correlators to remain sizable while
the MH quasiprobability has already become negative).

It is useful to contrast FTN with out-of-time-ordered correlators (OTOCs).
OTOCs are powerful diagnostics of operator growth and scrambling: they quantify how initially local operators fail to commute after time evolution and are
therefore sensitive to the spreading of quantum information across the system. Connections between OTOCs and quasiprobability distributions have been discussed previously in the KD framework~\cite{yunger2018quasiprobability,
arvidsson2024properties}. FTN addresses a related but distinct question. Rather than measuring the magnitude of operator growth or the decay of a four-point
correlator, it asks when a specified sequential measurement process first fails to admit a nonnegative quasiprobabilistic description. Thus, OTOCs characterize the growth of operator noncommutativity, whereas FTN identifies the onset of negativity in the KD and MH statistics associated with chosen local projectors.

The two diagnostics also differ operationally. Both depend on the chosen observables and on the state or ensemble in which they are evaluated. However, many OTOC protocols require time-reversal operations, echo sequences, ancilla-assisted interferometry, or randomized measurement schemes. By contrast, the KD and the MH quasiprobability underlying FTN can be reconstructed from sequential weak-then-strong measurements of local projectors~\cite{monroe2021weak,lostaglio2023kirkwood,wagner2024quantum}, without requiring full state tomography or explicit reversal of the many-body dynamics. This local-measurement structure is central to the present construction: the same framework applies to a single site or to spatially separated sites, thereby providing a site-resolved onset time for nonclassical sequential statistics.

\section{Results}
\label{Result}
We present results for both the integrable transverse-field Ising chain ($h_x=0$) and its nonintegrable extension with a finite longitudinal field ($h_x\neq 0$). 
The open boundary conditions break translational invariance, leading the FTN to vary with the spatial location of the measured observables. Unless stated otherwise, we focus on on-site measurements at a boundary site.
For such fixed local probes, $t_{\mathrm{FTN}}^{mm}$ depends only weakly on
the total chain length. Spatially separated measurements can behave
differently, since increasing $N$ may also increase the distance between the
two measured sites. The boundary modifies the local excitation and propagation pathways and can
therefore shift both the FTN and the corresponding QSL relative to bulk sites.
This position dependence is most pronounced in the interaction-dominated
regime. A full comparison is provided in App.~\ref{FTN_position}.

\begin{figure*}
\centering
\includegraphics[width =.48\linewidth, height=.35\linewidth]{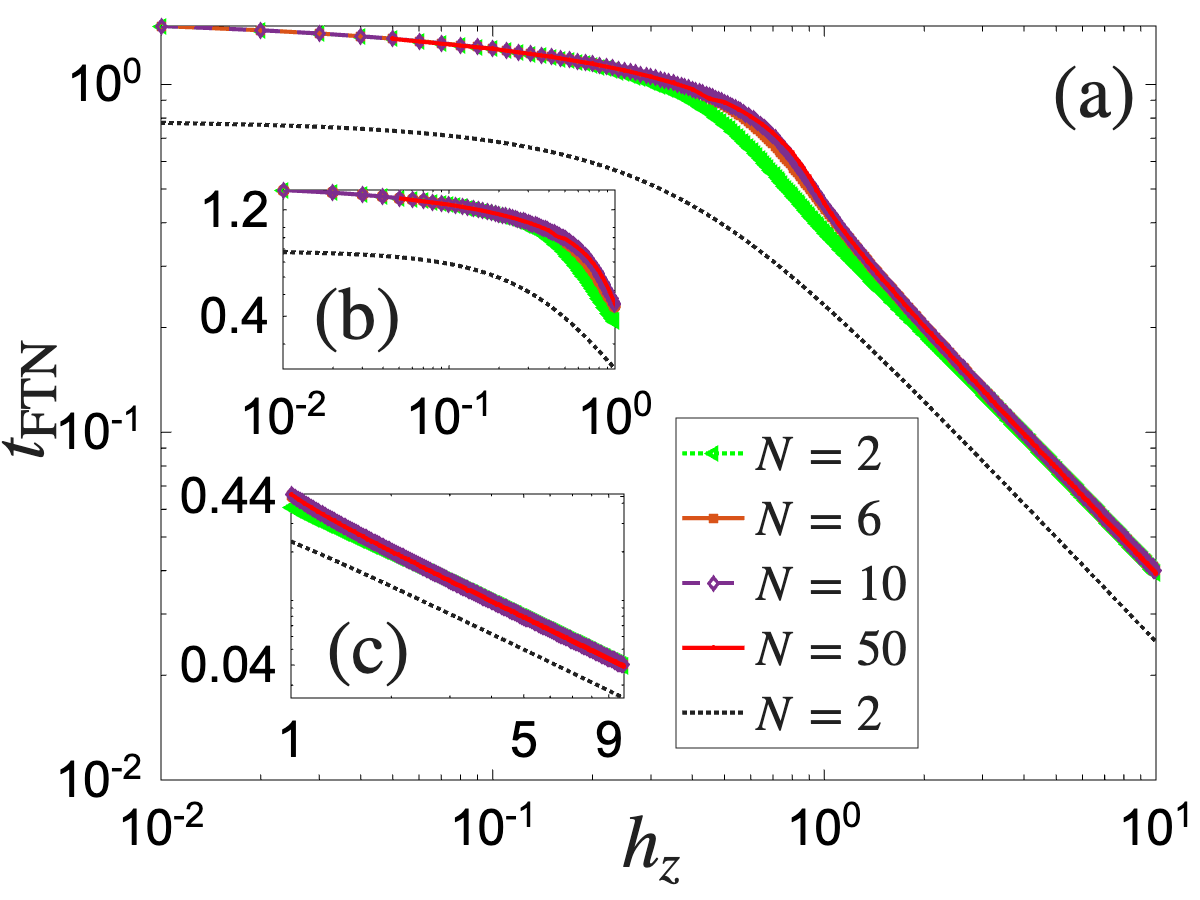} 
\includegraphics[width =.48\linewidth, height=.35\linewidth]{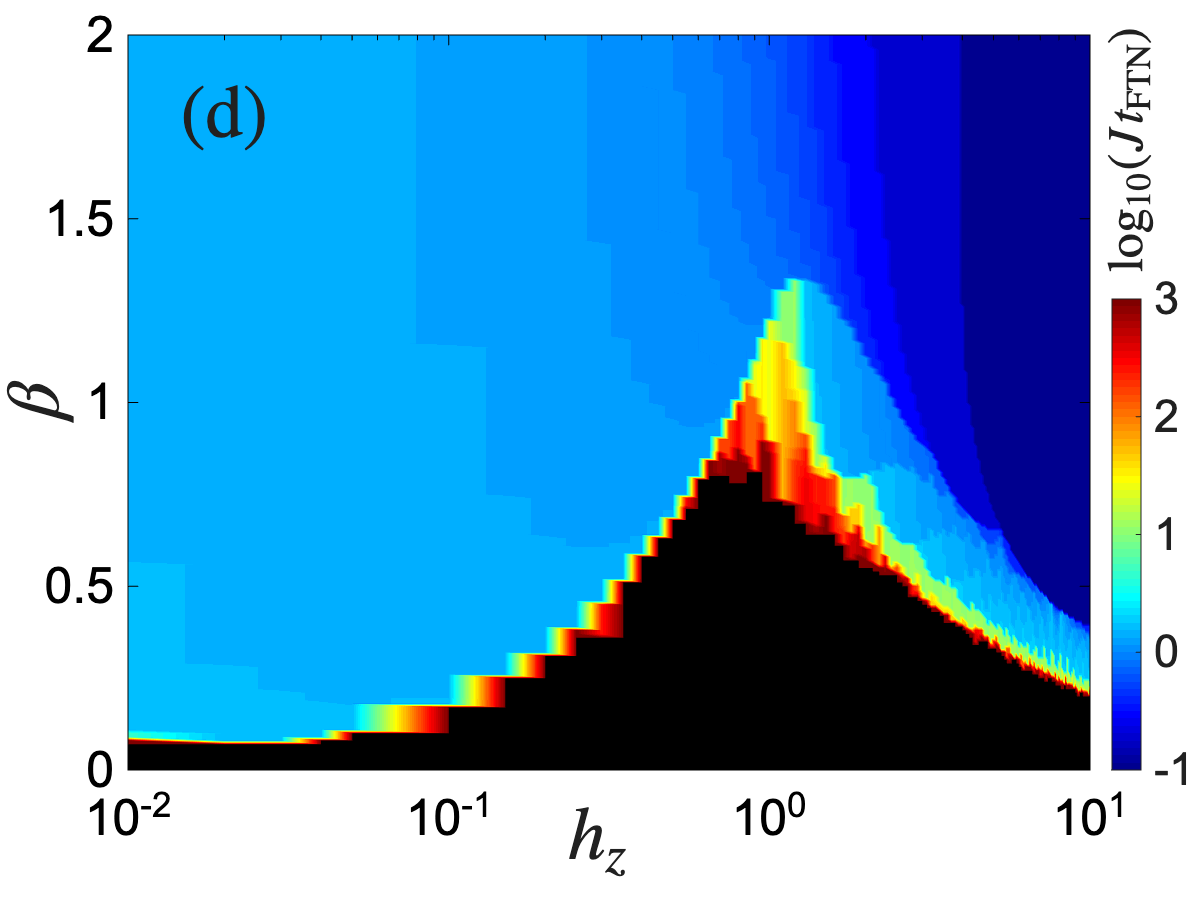} 
\caption{First-time negativity \(t_{\rm FTN}\) for the integrable transverse-field
Ising model \((h_x=0)\). All times are reported in units of \(1/J\).
(a) Zero-temperature \(t_{\rm FTN}\) as a function of \(h_z\) for
representative system sizes. The use of different line styles and markers
distinguishes the plotted values of \(N\); their near-overlap in the asymptotic
regimes shows the weak system-size dependence of the onset time. The black
dotted line indicates the corresponding QSL bound.
(b) Enlarged view of the weak-field regime. The numerical FTN approaches a
finite value as $h_z\to0^+$, with the square-root correction derived in
Eq.~\eqref{eq:tFTNsmalhz}. 
(c) Enlarged view of the strong-field regime, where the numerical FTN is
consistent with the asymptotic behavior $t_{\mathrm{FTN}}\propto1/h_z$
derived in Eq.~\eqref{eq:tFTNlargehz}. 
In panels (a)-(c), the black dotted curve denotes the corresponding QSL
bound.
(d) Finite-temperature map of \(\log_{10}(Jt_{\rm FTN})\) as a function of
\(h_z\) and inverse temperature \(\beta \), for \(N=8\). Black regions
indicate parameter values where no negativity was observed within the simulated
time window.}
\label{tl_hx_auto}
\end{figure*}

\subsection*{Transverse-field Ising model}

In the transverse-field Ising chain ($h_x=0$) with the system initially prepared in its ground state, we find that among all single-qubit observables only $\sigma_z$ exhibits quasiprobability negativity. The relevant quasiprobability is defined in terms of the local projectors $\Pi_0=\ketbra{0}{0}$ and $\Pi_1=\ketbra{1}{1}$ on site $m$ and their time evolution $\Pi_\gamma^m(t)$ as described above. These projectors form the underlying KD distribution from which physical observables can be reconstructed.

In particular, the local change in polarization along the transverse field reads as
\begin{equation}
\langle \sigma_z^m(t)\rangle - \langle \sigma_z^m(0)\rangle = 
\sum_{\gamma,\delta=0}^1 (\lambda_\gamma-\lambda_\delta)\,q_{\gamma\delta}^{mm}(t),
\end{equation}
while higher moments follow from analogous combinations of KD entries. Similarly, the two-time autocorrelation function of the transverse polarization can be expressed as
\begin{equation}
\Re C_{\sigma_z^m\sigma_z^m}(t) =\sum_{\gamma,\delta=0}^1 \lambda_\gamma \lambda_\delta\,q_{\gamma\delta}^{mm}(t),
\end{equation}
with eigenvalues $\lambda_0=1$ and $\lambda_1=-1$.

\subsubsection*{Zero temperature}

Figure~\ref{tl_hx_auto}(a) shows the FTN, $t_{\mathrm{FTN}}$, as a function of the transverse field $h_z$ for the integrable transverse-field Ising model ($h_x=0$) initialized in its ground state. 
The evolution time $t_{\mathrm{FTN}}$ marks the first moment when the MH quasiprobability of $\sigma_z^m$ acquires negative values. 
Its dependence on $h_z$ reveals three distinct dynamical regimes, separated by the quantum critical point $h_z=J$.

\paragraph*{Vanishing-negativity endpoints.}
There are two limiting cases in which no negativity is generated in the
\(\sigma_z\)-based MH quasiprobability considered here. For \(J=0\), the
Hamiltonian \(H=-h_z\sum_i\sigma_i^z\) commutes with the local
\(\sigma_z\) projectors, so that \(\Pi_\gamma^m(t)=\Pi_\gamma^m\). The MH
quasiprobability then reduces to an ordinary joint probability and remains
nonnegative.

The second endpoint is \(h_z=0\). In this case the absence of negativity is
seen most directly from the analytical quasiprobability. For example, the
two-site expression derived in App.~\ref{app:asymptotic_scale} gives
\begin{equation}
\Re q_{11}^{11}(t)\big|_{h_z=0}
=
\frac{1}{2}\cos^2(Jt)\ge 0 .
\end{equation}
Thus no finite first-time negativity exists at this isolated point;
equivalently, one may assign \(t_{\mathrm{FTN}}=\infty\) by convention when no
negative entry appears.

\paragraph*{Interaction-dominated ferromagnetic ordered regime 
\((0<h_z\ll J)\).}
For weak but finite transverse fields, the exchange interaction dominates and,
for \(J>0\), the system lies in the ferromagnetic ordered regime. The behavior
of the FTN in this regime can be understood as a perturbation away from the
nonnegative endpoint \(h_z=0\). The transverse field introduces a small
negative correction to the otherwise nonnegative quasiprobability, so the first
negative crossing occurs near the zeros of the unperturbed oscillatory factor.
As derived in App.~\ref{app:asymptotic_scale}, this gives
\begin{equation}
t_{\mathrm{FTN}} \approx
\frac{1}{J}\left(\frac{\pi}{2}-\sqrt{\frac{2h_z}{J}}\right),
\qquad 0<h_z\ll J .
\label{eq:tFTNsmalhz}
\end{equation}
This square-root decrease of \(t_{\mathrm{FTN}}\) with increasing transverse
field is consistent with the finite-\(h_z\) data shown in
Fig.~\ref{tl_hx_auto}(b). The endpoint \(h_z=0\) is special: the
quasiprobability remains nonnegative for all times, whereas for any finite
\(h_z>0\) a negative region appears. Hence the finite-\(h_z\) curve does not
diverge as \(h_z\to0^+\); instead, the onset time approaches a finite value
close to \(\pi/(2J)\), while the amount of negativity becomes arbitrarily
small.

\paragraph*{Field-dominated paramagnetic regime ($h_z\gg J$).}
In this regime, the transverse field sets the rapid oscillation frequency,
while the exchange interaction provides the noncommuting perturbation required
to generate negativity. The onset time decreases algebraically as
(see App.~\ref{app:asymptotic_scale})
\begin{equation}
t_{\mathrm{FTN}}
\approx
\frac{\pi h_z}{J^2+8h_z^2}
\simeq
\frac{\pi}{8h_z},
\qquad
h_z\gg J,
\label{eq:tFTNlargehz}
\end{equation}
consistent with the $1/h_z$ behavior shown in Fig.~\ref{tl_hx_auto}(c).
Thus, the transverse field determines the fast timescale of the first
crossing, whereas the exchange interaction controls the small deviation from
a nonnegative joint probability. Moreover, for $h_z\gg J$ the exchange term only weakly perturbs the nearly
$z$-polarized limit, so the MHQ entries deviate from an ordinary joint
probability by a parametrically small oscillation amplitude
$\sim J^{2}/h_z^{2}$ (see App.~\ref{app:asymptotic_scale}). Consequently, any
negative ``dip'' that appears becomes increasingly shallow (and the maximal
negativity scales down accordingly), making the onset progressively harder to
resolve and more susceptible to finite precision, decoherence, or thermal
mixing. In the strict $h_z\to\infty$ limit the quasiprobability becomes
nonnegative at all times.

The collapse of curves for system sizes $N=2$-$50$ indicates that $t^{mm}_{\mathrm{FTN}}$ is a local quantity, largely unaffected by system size (see App.~\ref{ana_calc} for the exact calculation for general system size $N$).  
The black dotted line in Fig.~\ref{tl_hx_auto}(a)-(c) represents the corresponding QSL bound, which provides a lower geometric bound for the time at which the KD distribution can become nonpositive.  
The bound exhibits a similar trend in the different regimes as $h_z$ varies and consistently lies below the numerically observed $t_{\mathrm{FTN}}$, reflecting that the bound captures the minimal geometric rate of state-projector incompatibility, whereas the actual onset of negativity depends on dynamical details of the many-body evolution.
Moreover, we note that the QSL bound, while informative about the maximal rate of change of expectation values, does not necessarily coincide with the emergence of nonclassicality: in cases where the underlying quasiprobability remains nonnegative at all times, the QSL still defines a finite evolution speed, but this motion occurs entirely within the classical domain of compatible observables (see App.~\ref{App:QSL}).

The behavior of the quasiprobability is independent of the sign of $J$ (see App.~\ref{FTN_J_s} for the proof). Consequently, the presence and magnitude of quasiprobability negativity are also insensitive to whether the interaction is ferromagnetic or antiferromagnetic in the integrable model.

\subsubsection*{Finite temperature}

We next examine how thermal fluctuations modify the onset of negativity. 
At finite temperature, the initial ground state is replaced by a Gibbs state,
\begin{equation}
\rho_0 = \frac{e^{-\beta H}}{Z}, \qquad 
Z = \mathrm{Tr}\!\left(e^{-\beta H}\right),
\end{equation}
with inverse temperature $\beta = 1/k_BT$ (setting $k_B = 1$). 
Figure~\ref{tl_hx_auto}(d) shows the resulting $t_{\mathrm{FTN}}$ landscape as a function of both $\beta$ and the transverse field $h_z$.

At low temperatures (large $\beta$), the behavior closely follows the zero-temperature limit: $t_{\mathrm{FTN}}$ decreases monotonically with increasing $h_z$, showing the familiar crossover between the interaction-dominated ($h_z\!\ll\!J$) and field-dominated ($h_z\!\gg\!J$) regimes.  
As temperature rises, however, thermal excitations begin to mask quantum coherence, reducing the interference required for the quasiprobability to become negative.  
This suppression is reflected by the growing black region in Fig.~\ref{tl_hx_auto}(d), where no negativity is detected.  
Physically, thermal mixing damps the off-diagonal components of $\rho_0$, weakening the overlap between the noncommuting projectors $\Pi_\gamma^m$ and $\Pi_\gamma^m(t)$ that give rise to negativity.

The sharp $T=0$ transition around $h_z = J$ broadens into a smooth finite-temperature crossover.  
In the $(h_z,\beta)$ plane this appears as a ridge of large
$t_{\mathrm{FTN}}$ in a band of fields around the zero–temperature critical
point, which is most pronounced at intermediate temperatures
($\beta \sim 1$) and gradually weakens as one moves away from criticality in
either $h_z$ or temperature. The growing black region at small $\beta$
indicates parameter values where no negativity is observed within our
simulation window; its boundary defines, for each $h_z$, a threshold
inverse temperature $\beta_c(h_z)$ below which the MH quasiprobability
remains nonnegative. The overall shape of this boundary can be interpreted as a
competition between the thermal time scale $1/T$ and the intrinsic dynamical
time scales of the chain, set by the many-body gap near $h_z\simeq J$ and by the rapid
field-induced oscillation scale at large $h_z$.

Far from criticality, the relevant dynamical scale is no longer dictated by the
many-body gap but by the rapid transverse oscillations induced by the strong
field. In the limit $h_z \gg J$, the dominant field aligns the spins nearly
along the $z$-axis, while the weaker exchange term $J$ drives small transverse
fluctuations that oscillate at a rate set by $h_z$. These fast oscillations
limit the buildup of coherent interference responsible for the emergence of
negativity, leading to a characteristic timescale $t_{\mathrm{FTN}} \propto
1/h_z$. As temperature increases, thermal averaging over these rapid
oscillations becomes effective once $k_B T$ is comparable to $h_z$, further
suppressing the appearance of negativity even when the excitation gap remains
large. Thus, both mechanisms, the thermal smearing of correlations near the
critical region and the field-induced suppression of coherence at large
$h_z$, jointly determine the shape and extent of the finite-temperature
boundary in Fig.~\ref{tl_hx_auto}(d).

In the high-temperature limit, the Gibbs state becomes fully mixed,
\begin{equation}
\rho_{\beta\to0}
=
\frac{\mathbb{I}}{\mathcal{D}},
\qquad \mathcal{D}=2^N .
\end{equation}
Substitution into the MH quasiprobability gives
\begin{equation}
q_{\gamma\delta}^{mn}(t;\beta=0)
=
\frac{1}{\mathcal{D}}
\Re\,\mathrm{Tr}\!\left[
\Pi_\gamma^m(t)\Xi_\delta^n
\right]
=
\frac{1}{\mathcal{D}}
\mathrm{Tr}\!\left[
\Xi_\delta^n\Pi_\gamma^m(t)\Xi_\delta^n
\right]
\ge 0.
\end{equation}
Here we used the cyclicity of the trace and \((\Xi_\delta^n)^2=\Xi_\delta^n\).
The final operator,
\(\Xi_\delta^n\Pi_\gamma^m(t)\Xi_\delta^n\), is positive semidefinite because
\(\Pi_\gamma^m(t)\) is a projector. Thus, at infinite temperature, no negative
MH entry can appear and no finite \(t_{\mathrm{FTN}}\) exists.

For small but finite \(\beta\), the Gibbs state can be expanded as
\begin{equation}
\rho_\beta
=
\frac{e^{-\beta H}}{\mathrm{Tr}(e^{-\beta H})}
=
\frac{\mathbb{I}}{\mathcal{D}}
-
\frac{\beta}{\mathcal{D}}H
+
\mathcal{O}(\beta^2),
\end{equation}
where we used \(\mathrm{Tr}H=0\) for the Ising Hamiltonian in
Eq.~\eqref{eq:H}. Hence
\begin{equation}
q_{\gamma\delta}^{mn}(t;\beta)
=
q_{\gamma\delta}^{mn}(t;0)
-
\frac{\beta}{\mathcal{D}}
\Re\,\mathrm{Tr}\!\left[
\Pi_\gamma^m(t)\Xi_\delta^n H
\right]
+
\mathcal{O}(\beta^2).
\end{equation}
Therefore, high-temperature negativity can only arise from finite-\(\beta\)
corrections to the fully mixed state. This explains the suppression and
eventual disappearance of FTN as temperature increases, as shown in
Fig.~\ref{tl_hx_auto}(d).

\subsection*{Spatio-temporal negativity at finite separation}
\label{sec:spatial}

So far, we have considered local quasiprobabilities with $m=n$. We now place
the two $\sigma_z$ projectors at different lattice sites and study
$q_{\gamma\delta}^{mn}(t)$ for the ordered measurement sequence
``measure $\sigma_z^n$ at $t=0$, and then measure $\sigma_z^m$ at time
$t$.'' We fix the time-evolved projector $\Pi_\gamma^m(t)$ at the boundary site
$m=1$ and place the projector $\Xi_\delta^n$ at site $n=1+d$, where
$d=|n-m|$ is the separation between the two sites.

For the local projectors
$\Pi_\gamma^m=(\mathbb{I}+\lambda_\gamma\sigma_z^m)/2$ and
$\Xi_\delta^n=(\mathbb{I}+\lambda_\delta\sigma_z^n)/2$, with
$\lambda_0=1$ and $\lambda_1=-1$, the MH quasiprobability can be written as
\begin{equation}
q_{\gamma\delta}^{mn}(t)
=
\frac{1}{4}
\left[
1
+\lambda_\gamma\langle\sigma_z^m\rangle
+\lambda_\delta\langle\sigma_z^n\rangle
+\lambda_\gamma\lambda_\delta
\operatorname{Re}
C_{\sigma_z^m\sigma_z^n}(t)
\right],
\label{eq:spatial_MH}
\end{equation}
where
\begin{equation}
C_{\sigma_z^m\sigma_z^n}(t)
=
\left\langle
\sigma_z^m(t)\sigma_z^n(0)
\right\rangle .
\label{eq:spatial_correlator}
\end{equation}
The initial state is the ground state of the same Hamiltonian that generates
the time evolution. Therefore,
$\langle\sigma_z^m(t)\rangle=\langle\sigma_z^m\rangle$, and all time
dependence in Eq.~\eqref{eq:spatial_MH} is contained in the two-time
correlator.

Equation~\eqref{eq:spatial_MH} provides a direct threshold condition for
negativity:
\begin{equation}
\lambda_\gamma\lambda_\delta
\operatorname{Re}
C_{\sigma_z^m\sigma_z^n}(t)
<
-
\left[
1
+\lambda_\gamma\langle\sigma_z^m\rangle
+\lambda_\delta\langle\sigma_z^n\rangle
\right].
\label{eq:spatial_negativity_threshold}
\end{equation}
Thus, correlations or noncommutativity alone do not guarantee a negative MH
entry. The real part of the correlator must cross a finite threshold determined
by the local polarizations and the chosen pair of projectors.

\begin{figure*}
\centering
\includegraphics[width =.48\linewidth, height=.35\linewidth]{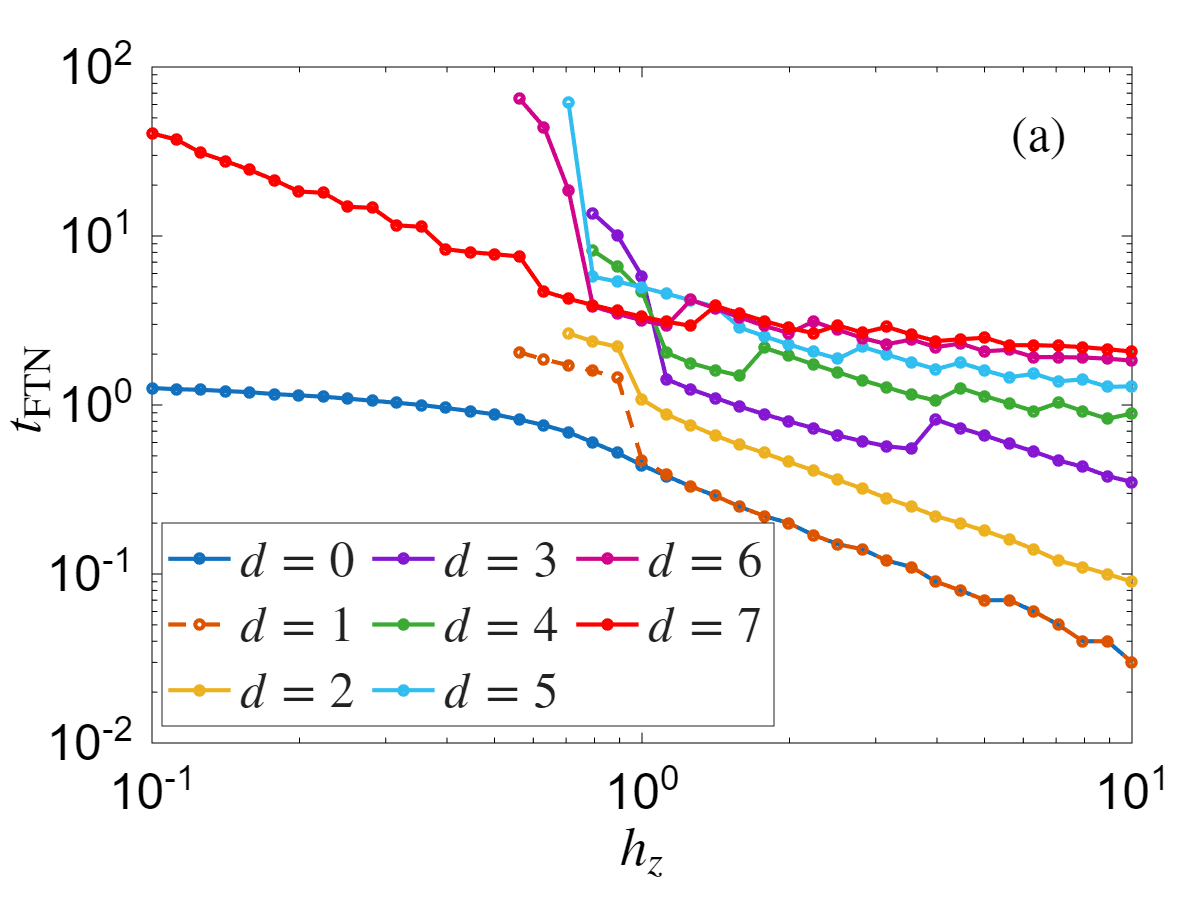} 
\includegraphics[width =.48\linewidth, height=.35\linewidth]{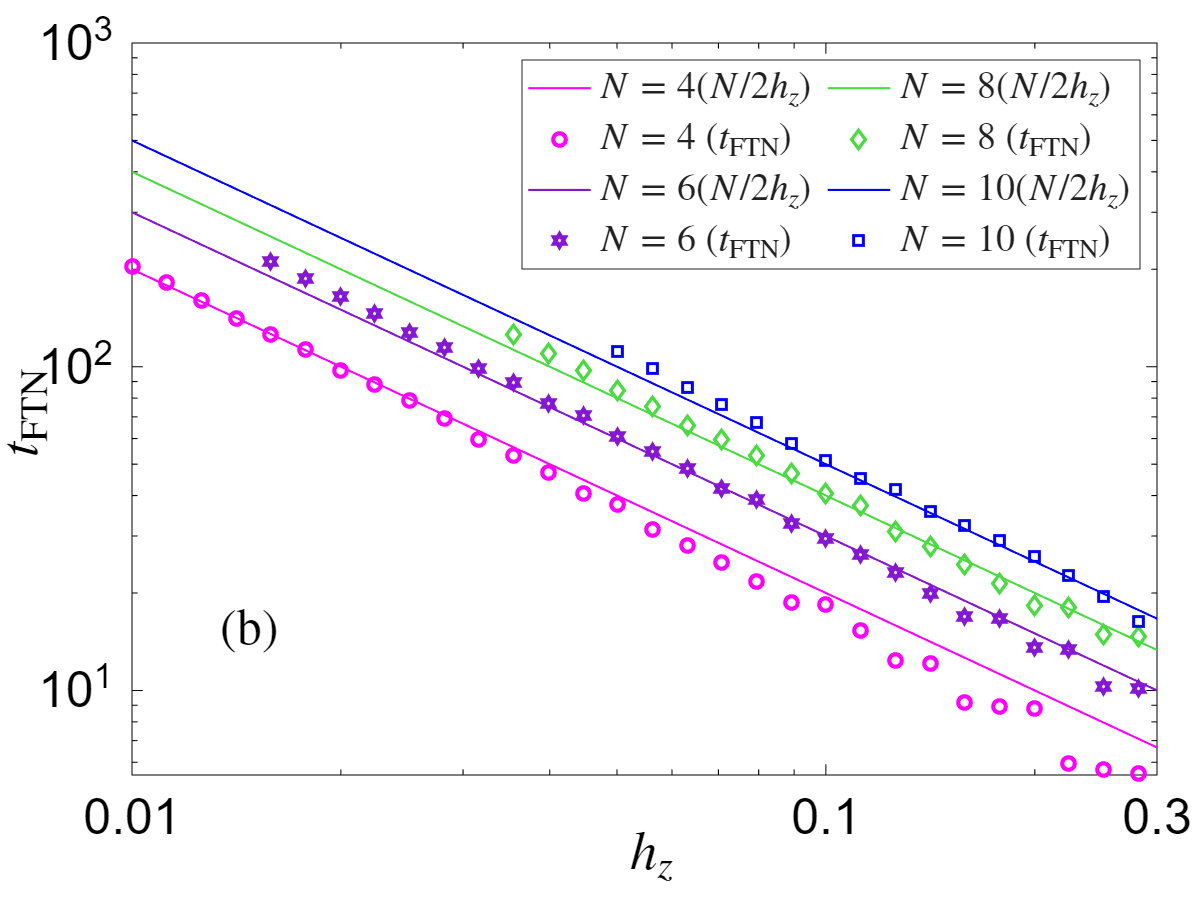} 
\caption{Spatio-temporal dependence of the first-time negativity in the integrable
transverse-field Ising model with open boundary conditions, $J=1$, $h_x=0$,
and the ground state as the initial state.
(a) $t_{\mathrm{FTN}}^{1,1+d}$ as a function of the transverse field $h_z$
for a chain of $N=8$ sites. The time-evolved projector $\Pi_\gamma^1(t)$ is fixed at the boundary site
$m=1$, while $\Xi_\delta^{1+d}$ is located at $n=1+d$, with
$d=0,\ldots,7$. The on-site result corresponds to $d=0$, while $d=7$
corresponds to measurements at the two opposite boundaries. For generic
separated sites, no negative MH entry is observed in part of the weak-field
regime within the numerical observation interval, and a finite FTN appears
abruptly as $h_z$ approaches the interaction scale. The opposite-boundary
configuration exhibits a distinct weak-field branch.
(b) Opposite-boundary first-time negativity,
$t_{\mathrm{FTN}}^{1N}$, for $N=4,6,8,$ and $10$ in the weak-field regime.
Symbols show the numerical FTN, while the solid lines show the ballistic
reference scale $N/(2h_z)$. The approximately parallel curves on logarithmic
axes and their systematic increase with $N$ are consistent with
$t_{\mathrm{FTN}}^{1N}\propto N/h_z$.  All times are given in units of $1/J$.}
\label{tl_int_d}
\end{figure*}

Figure~\ref{tl_int_d}(a) shows $t_{\mathrm{FTN}}^{1,1+d}$ as a function of
$h_z$ for an open chain with $N=8$, $J=1$, and $h_x=0$. The on-site result,
$d=0$, varies smoothly with the transverse field. For the separated-site
configurations $1\leq d\leq N-2$, no negative MH entry is found in the
weak-field regime within the numerical observation interval. A finite
$t_{\mathrm{FTN}}^{1,1+d}$ then appears abruptly as $h_z$ approaches the
interaction scale.

The missing weak-field portions of these curves do not represent very large
measured values of $t_{\mathrm{FTN}}^{1,1+d}$. Rather, none of the MH entries
crosses zero before the maximal simulation time. The field at which a finite
FTN first appears depends on the separation and marks a dynamical onset within
the observation interval. It should not be interpreted as an additional
equilibrium phase transition. The smaller kinks and nonmonotonic features
reflect the first-crossing definition of FTN, since different oscillatory
extrema can determine the earliest threshold crossing as $h_z$ is varied.

A qualitatively different branch appears for $d=N-1$, where the measurements
are located at the two opposite boundaries. Reflection symmetry gives
$
\langle\sigma_z^1\rangle
=
\langle\sigma_z^N\rangle 
$
and for $h_z>0$, the local polarization is nonnegative. Using
$-1\leq\operatorname{Re}C_{\sigma_z^1\sigma_z^N}(t)\leq1$, one finds that
$q_{00}^{1N}$, $q_{01}^{1N}$, and $q_{10}^{1N}$ remain nonnegative. The
entry that can become negative is
\begin{equation}
q_{11}^{1N}(t)
=
\frac{1}{4}
\left[
1
-2\langle\sigma_z^1\rangle
+\operatorname{Re}
C_{\sigma_z^1\sigma_z^N}(t)
\right],
\label{eq:q11_boundaries}
\end{equation}
with the threshold
\begin{equation}
\operatorname{Re}
C_{\sigma_z^1\sigma_z^N}(t)
<
2\langle\sigma_z^1\rangle-1.
\label{eq:boundary_negativity_threshold}
\end{equation}
At $h_z=0$, $\langle\sigma_z^1\rangle=0$, so negativity would require the
correlator to become smaller than $-1$, which is impossible. A finite
transverse field shifts the threshold above $-1$ and allows the
opposite-boundary correlator to generate a negative MH entry.

The opposite-boundary geometry is also distinguished by the excitations
selected by the two projectors. In the ordered regime, a boundary
$\sigma_z$ operator couples predominantly to the one-domain-wall sector,
whereas a bulk $\sigma_z$ operator couples at leading order to configurations
containing two domain walls. The two boundary operators therefore share a
leading propagation channel that is absent for a generic boundary-bulk pair.
A more detailed threshold and domain-wall analysis is given in
App.~\ref{app:spatial_thresholds}.

The exact TFIM quasiparticle spectrum gives the maximal group velocity
\begin{equation}
v_{\max}
=
2\min(J,h_z),
\label{eq:max_group_velocity_spatial}
\end{equation}
for $h_z\geq0$~\cite{pfeuty1970one}. In the ordered regime, $h_z<J$, the
corresponding boundary-to-boundary propagation scale is
\begin{equation}
t_{\mathrm{prop}}
\sim
\frac{N-1}{2h_z}.
\label{eq:boundary_propagation_scale}
\end{equation}
This finite propagation scale is consistent with the general locality
structure of short-range quantum spin systems~\cite{lieb1972finite}.

Figure~\ref{tl_int_d}(b) shows $t_{\mathrm{FTN}}^{1N}$ for several system
sizes. The weak-field curves are approximately straight and parallel on
logarithmic axes, indicating
\begin{equation}
t_{\mathrm{FTN}}^{1N}
\propto
\frac{N}{h_z},
\qquad
h_z<J.
\label{eq:boundary_FTN_scaling}
\end{equation}
Their numerical values are close to $N/(2h_z)$ over the field range shown.
This agreement is consistent with the boundary-to-boundary FTN being governed
by the ballistic propagation scale. The FTN need not coincide exactly with
$t_{\mathrm{prop}}$, because negativity appears only after the end-to-end
correlator crosses the finite threshold in
Eq.~\eqref{eq:boundary_negativity_threshold}.
In the field-dominated regime, $h_z\gg J$, the maximal propagation velocity
saturates at $2J$. Increasing $h_z$ therefore no longer accelerates the
propagation of correlations between distant sites. Consistent with this
behavior, the FTN curves at larger separations become only weakly dependent on
$h_z$ and retain a distance-dependent scale governed primarily by $J$.
Short-distance curves can retain a stronger field dependence because they
combine local oscillations on the scale $1/h_z$ with only a short propagation
distance.

The spatial results therefore distinguish the local onset from the
propagation-controlled boundary-to-boundary onset. The on-site
$t_{\mathrm{FTN}}^{mm}$ depends only weakly on the total system size, whereas
$t_{\mathrm{FTN}}^{1N}$ contains a traversal timescale that grows with $N$ in
the ordered regime.


%

\subsection*{Nonintegrable case: longitudinal field}
\begin{figure}
    \centering
    \includegraphics[width =.35\linewidth]{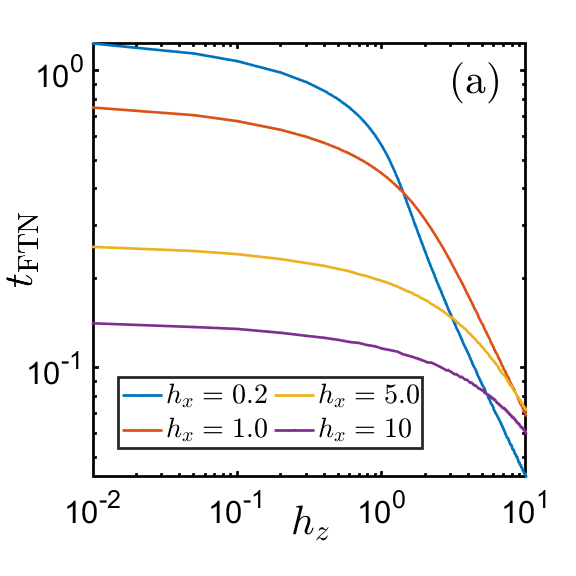}
   \includegraphics[width =.35\linewidth]{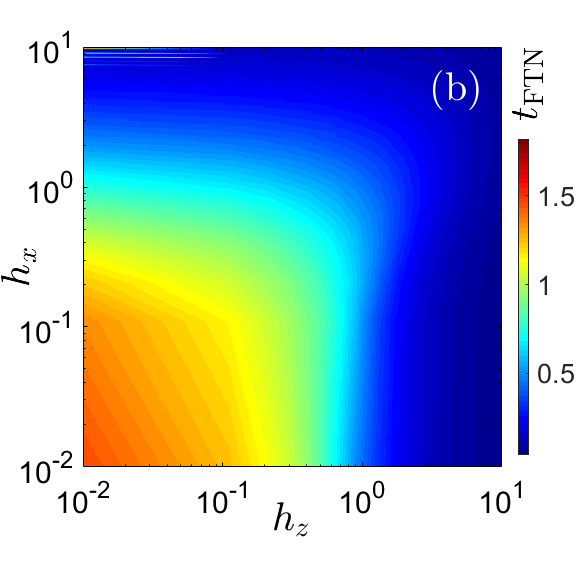}
    \includegraphics[width =.35\linewidth]{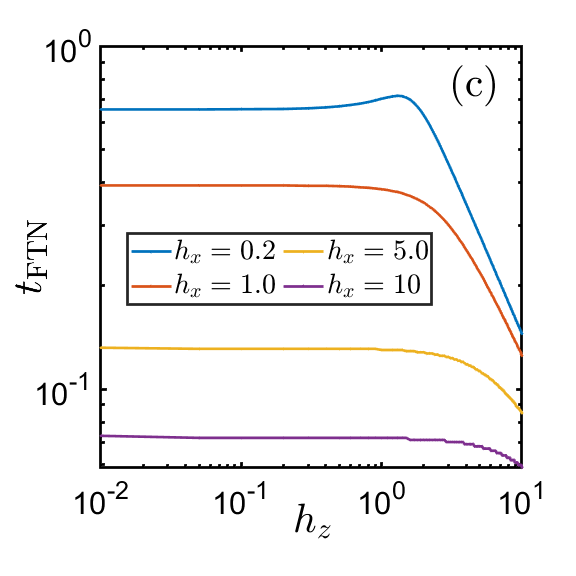}
      \includegraphics[width =.35\linewidth]{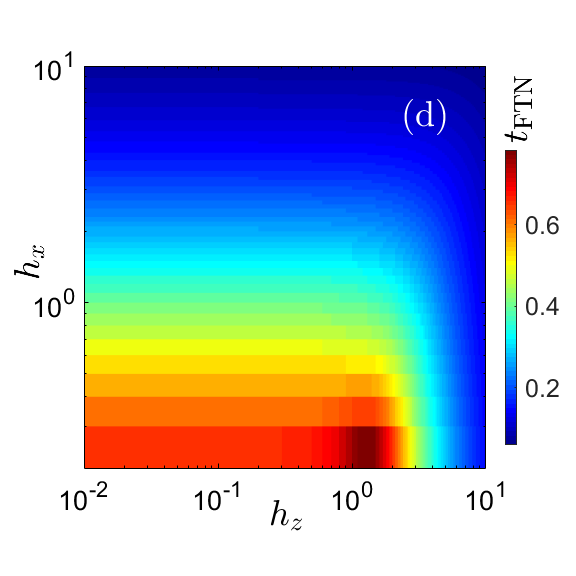} 
\caption{
First-time negativity $t_{\rm FTN}$ (in units of $1/J$) in the nonintegrable Ising chain with a longitudinal field $h_x$.  
(a,c) $t_{\rm FTN}$ as a function of the transverse field $h_z$ for several fixed values of $h_x$ (see legends).  
(b,d) $t_{\rm FTN}$ as a function of both $h_z$ and $h_x$.  
In panels (a,b) the local observable is $V=W=\sigma_z^1$, while in panels (c,d) it is $V=W=\sigma_x^1$.  
In all cases $N=8$ and $J=1$.
}\label{tl_hx_3D}
\end{figure}

To explore the effect of breaking integrability, we now introduce a longitudinal field $h_x$ into the Ising Hamiltonian.  
This term explicitly breaks the $\mathbb{Z}_2$ spin-flip symmetry of the transverse-field model~\cite{heyl2018dynamical}, allowing additional local channels for generating quantum interference and hence quasiprobability negativity.  
Figures~\ref{tl_hx_3D}(a)-(d) summarize the dependence of the FTN on both $h_x$ and $h_z$ for two types of local projectors:  
panels (a,b) correspond to $\sigma_z$ and panels (c,d) to $\sigma_x$ observables.  
Unlike the integrable case, where the KD quasiprobability of $\sigma_x$ remained strictly nonnegative, here both observables exhibit finite negativity once $h_x$ is switched on.

For the $\sigma_z$ projectors, Figs.~\ref{tl_hx_3D}(a),(b), the longitudinal
field $h_x$ preserves the overall qualitative shape of $t_{\mathrm{FTN}}(h_z)$
but shifts it to shorter times. From the quasiprobability viewpoint, negativity
requires coherent superpositions between the two $\sigma_z$ eigenstates of the
local spin. When $h_x=0$, this mixing is generated only indirectly by the
exchange term, so a boundary $\sigma_z^m$ remains relatively protected and
$t_{\mathrm{FTN}}$ is large, especially at small $h_z$. Turning on $h_x$ adds a
local $\sigma_x$ term that does not commute with $\sigma_z^m$, creating such
superpositions much more efficiently and thereby shortening $t_{\mathrm{FTN}}$,
as seen by the downward shift of the curves in
Fig.~\ref{tl_hx_3D}(a) and the shrinking red region in the 2D map
Fig.~\ref{tl_hx_3D}(b). For $h_z \gg J$, the strong transverse field dominates
the dynamics, and all curves collapse onto a common $t_{\mathrm{FTN}}\propto
1/h_z$ tail with only weak residual dependence on $h_x$.

For the $\sigma_x$ projectors, Figs.~\ref{tl_hx_3D}(c),(d), the introduction of $h_x$ fundamentally changes the picture.  
At $h_x=0$, the $\mathbb{Z}_2$ symmetry protects $\sigma_x$ from generating negativity, leading to a purely classical KD distribution.  
Even a weak longitudinal field breaks this protection, allowing interference between parity sectors and yielding finite negativity.  
At small $h_x$, $t_{\mathrm{FTN}}$ is long and shows a shallow peak near $h_z\simeq J$, marking the transition between the interaction- and field-dominated regimes.  
As $h_x$ increases, this feature disappears and $t_{\mathrm{FTN}}$ shortens throughout the entire range of $h_z$, reflecting a faster spread of operator incompatibility induced by the longitudinal perturbation.  
The color map in Fig.~\ref{tl_hx_3D}(d) confirms this trend: for a fixed $h_z$, larger $h_x$ consistently leads to smaller $t_{\mathrm{FTN}}$ values.  
In the limit of large transverse fields, both $\sigma_z$ and $\sigma_x$ observables converge to the same asymptotic scaling $t_{\mathrm{FTN}}\!\sim\!1/h_z$, since the fast transverse rotations dominate over exchange and longitudinal contributions.

In the strong-field regime ($h_z\gg |J|$), $t_{\mathrm{FTN}}$ is controlled
primarily by the transverse-field timescale and is therefore only weakly
sensitive to the sign of $J$.  Differences between ferromagnetic and
antiferromagnetic couplings emerge mainly at low fields, where the exchange and
longitudinal terms compete more directly with the transverse dynamics (see App.~\ref{FTN_J_s} for details).

\section{Conclusions}
\label{Conclusion}

We introduced the first-time negativity of the Margenau-Hill quasiprobability,
$t_{\mathrm{FTN}}$, as an operational timescale for the emergence of
quasiprobability negativity under sequential local measurements in many-body
dynamics. In the integrable transverse-field Ising model ($h_x=0$), we showed
that negativity emerges only for $\sigma_z$ probes and that $t_{\mathrm{FTN}}$
exhibits two dynamical regimes separated by a sharp crossover around the
quantum critical point $h_z=J$, governed respectively by interaction-dominated
and field-dominated physics. 
The weak system-size dependence of the on-site FTN highlights the local character of the single-site onset,
while increasing the temperature broadens the sharp zero-temperature feature into a finite-temperature crossover, and can
eventually wash out negativity altogether.

Comparison with the recently derived QSL for KD quasiprobabilities shows that the bound constrains only the maximal kinematic rate of change, and can remain
finite even when the MH quasiprobability stays nonnegative at all times. By
contrast, $t_{\mathrm{FTN}}$ captures the actual onset of nonclassicality by
directly identifying when negativity first appears.

For spatially separated measurements in the integrable TFIM, the FTN reveals
two distinct behaviors. For generic separated sites, no negative MH entry is
observed over part of the weak-field regime within the accessible numerical
time window, and a finite FTN appears abruptly as the transverse field
approaches the interaction scale.
Measurements at the two opposite open boundaries exhibit a distinct
weak-field branch whose onset time grows approximately linearly with chain
length and inversely with the transverse-field strength. This behavior is
consistent with a ballistic boundary-to-boundary propagation timescale. In
the field-dominated regime, the onset at distant sites becomes only weakly
dependent on the transverse field and retains a pronounced dependence on
separation.

Breaking integrability with a longitudinal field $h_x$ lifts the symmetry
protection that kept $\sigma_x$ classical and opens new channels for
generating superpositions. As a result, both $\sigma_z$ and $\sigma_x$
observables develop negativity, and $t_{\mathrm{FTN}}$ retains robust
features, most notably a $1/h_z$ tail at strong fields, while reflecting the modified dynamics of the nonintegrable model.

Finally, our conclusions are largely insensitive to the sign of the exchange.
In the integrable case, the quasiprobability (and thus the presence and
magnitude of its negativity) is independent of $\mathrm{sign}(J)$. In the
nonintegrable case, $\mathrm{sign}(J)$ effects are weak at strong fields
($h_z\gg |J|$) and become noticeable mainly at low fields, where exchange
competes more directly with the longitudinal term.

These results show that quasiprobability negativity, and in particular the
onset time $t_{\mathrm{FTN}}$, provides an experimentally accessible probe of
dynamical quantum behavior beyond correlations alone. The diagnostic is robust
at the qualitative level because it is not tied to a fine-tuned parameter
point: in the examples studied here, $t_{\mathrm{FTN}}$ changes systematically
with field strength, temperature, spatial separation, and integrability
breaking. In practice, experimental uncertainties can be incorporated by
defining a thresholded onset time,
\begin{equation}
t_{\mathrm{FTN}}^{(\epsilon)}
=
\min\left\{
t>0:
\min_{\gamma,\delta} q_{\gamma\delta}^{mn}(t)<-\epsilon
\right\},
\end{equation}
where $\epsilon$ is set by the statistical and systematic uncertainty of the
measurement. This avoids assigning significance to unresolved zero crossings or
very shallow negative dips.

Beyond its diagnostic role, FTN also suggests a natural control
objective: one may seek to advance, delay, enhance, or suppress the emergence
of negativity through Hamiltonian engineering, state preparation, external
driving, or measurement design. The definition of FTN extends naturally to
higher-dimensional systems and multitime sequences whenever the corresponding
KD or MH quasiprobability can be reconstructed. Its practical use in such extended
settings will require addressing additional issues, including the choice of
measurement projectors, sampling cost, and finite-size or boundary effects.
The connection to information propagation is motivated by previous work
relating KD quasiprobabilities to scrambling and out-of-time-ordered
correlators~\cite{yunger2018quasiprobability,arvidsson2024properties}.
Experimentally, FTN is most directly relevant to platforms implementing
weak-then-strong measurement protocols, such as trapped ions, superconducting
qubits, Rydberg-atom arrays, and solid-state spin systems. It is also naturally
suited to gate-based quantum-computing platforms, where weak measurements can
be implemented through ancilla-assisted circuits or controlled weak couplings,
followed by projective readout and repeated sampling. Thus FTN is relevant both
to analog quantum simulators and to digital quantum processors, provided that
coherence times and measurement precision are sufficient to resolve the first
negative MH entry.

\section*{ACKNOWLEDGMENTS}
We thank Jonathan Ruhman for fruitful discussions and useful comments on the manuscript.

This research was supported by the ISF Grants No.~1364/21 and No.~3105/23 and Grant No.~2022312 from the United States-Israel Binational Science Foundation (BSF).



\section*{Data Availability}
The data supporting the findings of this study are openly available at \cite{github_data_FTN}.

\bibliography{OTOC_LRT}
\newpage

\appendix
\section{ Quantum speed limit for nonpositivity of the
KDQ}
\label{App:QSL}

The real part of the KD quasiprobability can be expressed as an expectation
value of a Hermitian operator. Defining
\begin{equation}
\rho_\delta^n
=
\frac{1}{2}
\left\{
\rho_0,\Xi_\delta^n
\right\},
\label{eq:QSL_rho_delta}
\end{equation}
we obtain
\begin{equation}
q_{\gamma\delta}^{mn}(t)
=
\operatorname{Re}
\operatorname{Tr}
\left[
\Pi_\gamma^m(t)\Xi_\delta^n\rho_0
\right]
=
\operatorname{Tr}
\left[
\rho_\delta^n\Pi_\gamma^m(t)
\right].
\label{eq:QSL_expectation}
\end{equation}
The operator $\rho_\delta^n$ is Hermitian, but it is not necessarily positive
semidefinite.

The local projector $\Pi_\gamma^m$ has rank and trace
$r_\gamma=\mathcal{D}/2$, where $\mathcal{D}=2^N$ is the Hilbert-space
dimension. To apply the QSL of Ref.~\cite{pratapsi2025quantum}, we introduce
the normalized projector
\begin{equation}
\widetilde{\Pi}_\gamma^m(t)
=
\frac{\Pi_\gamma^m(t)}{r_\gamma}
\label{eq:QSL_normalized_projector}
\end{equation}
and the corresponding normalized quasiprobability
\begin{equation}
\widetilde q_{\gamma\delta}^{mn}(t)
=
\operatorname{Tr}
\left[
\rho_\delta^n\widetilde{\Pi}_\gamma^m(t)
\right]
=
\frac{
q_{\gamma\delta}^{mn}(t)
}{
r_\gamma
}.
\label{eq:QSL_normalized_q}
\end{equation}
Since $r_\gamma>0$, the normalized and unnormalized quasiprobabilities cross
zero at exactly the same time.

\subsection*{Speed bound}

The symmetric logarithmic derivative $L_\gamma^m(t)$ associated with
$\widetilde{\Pi}_\gamma^m(t)$ is defined implicitly by
\begin{equation}
\frac{d}{dt}
\widetilde{\Pi}_\gamma^m(t)
=
\frac{1}{2}
\left\{
\widetilde{\Pi}_\gamma^m(t),
L_\gamma^m(t)
\right\}.
\label{eq:QSL_SLD_definition}
\end{equation}
Its variance is
\begin{equation}
\left(
\Delta L_\gamma^m
\right)^2
=
\operatorname{Tr}
\left[
\widetilde{\Pi}_\gamma^m(t)
\left(
L_\gamma^m(t)
\right)^2
\right]
-
\left[
\operatorname{Tr}
\left(
\widetilde{\Pi}_\gamma^m(t)
L_\gamma^m(t)
\right)
\right]^2.
\label{eq:QSL_SLD_variance}
\end{equation}
The second term vanishes because
$\operatorname{Tr}[\widetilde{\Pi}_\gamma^m(t)L_\gamma^m(t)]=0$.
For a time-independent Hamiltonian,
$\Delta L_\gamma^m$ is independent of time.

Let $x_{\min}$ and $x_{\max}$ denote the minimum and maximum eigenvalues of a
Hermitian operator $X$. The interpolation function entering the QSL is
\begin{equation}
E(X,\tau)
=
\begin{cases}
x_{\max},
&
\tau\leq0,
\\[1mm]
x_{\max}
\cos^2\left(\dfrac{\tau}{2}\right)
+
x_{\min}
\sin^2\left(\dfrac{\tau}{2}\right),
&
0\leq\tau\leq\pi,
\\[1mm]
x_{\min},
&
\tau\geq\pi.
\end{cases}
\label{eq:QSL_interpolation}
\end{equation}
For $x_{\min}\leq x\leq x_{\max}$, its inverse is
\begin{equation}
\tau(X,x)
=
\arccos
\left[
\frac{
2x-x_{\min}-x_{\max}
}{
x_{\max}-x_{\min}
}
\right],
\label{eq:QSL_interpolation_angle}
\end{equation}
which satisfies
\begin{equation}
E\!\left[X,\tau(X,x)\right]=x.
\end{equation}

In the present application, $X=\rho_\delta^n$. Therefore, $x_{\min}$ and
$x_{\max}$ in Eqs.~\eqref{eq:QSL_interpolation} and
\eqref{eq:QSL_interpolation_angle} are the extremal eigenvalues of
$\rho_\delta^n$, not those of $\Pi_\gamma^m$.

The QSL derived in Ref.~\cite{pratapsi2025quantum} gives the lower bound
\begin{equation}
\widetilde q_{\gamma\delta}^{mn}(t)
\geq
E
\left[
\rho_\delta^n,
\tau_{\gamma\delta;mn}^{(0)}
+
\Delta L_\gamma^m t
\right],
\label{eq:QSL_lower_bound}
\end{equation}
where
\begin{equation}
\tau_{\gamma\delta;mn}^{(0)}
=
\tau
\left[
\rho_\delta^n,
\widetilde q_{\gamma\delta}^{mn}(0)
\right].
\label{eq:QSL_initial_angle}
\end{equation}

Provided that zero lies within the spectral interval of $\rho_\delta^n$, the
earliest time at which the lower bound can reach zero is
\begin{equation}
T_{\gamma\delta;mn}^{\mathrm{re}}
=
\frac{
\tau\left(\rho_\delta^n,0\right)
-
\tau_{\gamma\delta;mn}^{(0)}
}{
\Delta L_\gamma^m
}.
\label{eq:QSL_time_nonpositivity}
\end{equation}
This is a lower bound on the time at which the corresponding MH entry can
become nonpositive. A finite
$T_{\gamma\delta;mn}^{\mathrm{re}}$ does not imply that the actual
quasiprobability reaches zero or becomes negative.

If zero does not belong to the interval
$[x_{\min},x_{\max}]$, the expectation value cannot reach zero and no finite
time-to-nonpositivity bound exists. If
$\widetilde q_{\gamma\delta}^{mn}(0)\leq0$, the corresponding onset time is
already zero.

\begin{figure}
    \centering
\includegraphics[width=0.49\linewidth,height=0.35\linewidth]{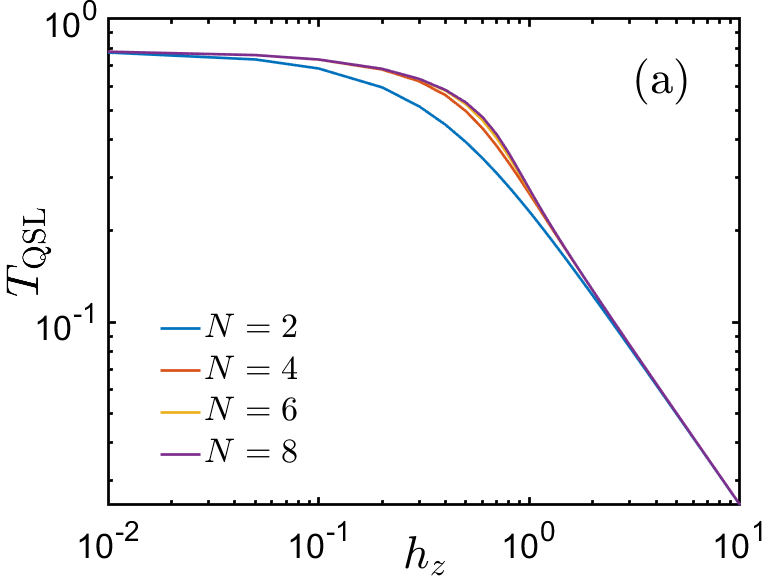}
\includegraphics[width=0.49\linewidth,height=0.35\linewidth]{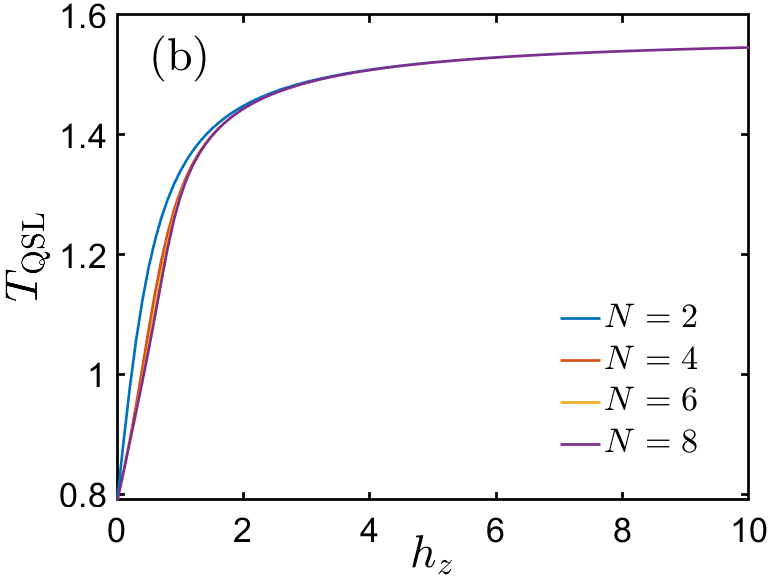}
\caption{
Quantum speed-limit time for different local projector combinations in the integrable transverse-field Ising model.  
Shown is the time to possible nonpositivity of the real part of the KD quasiprobability, $T_{\rm QSL}$, as a function of the transverse field $h_z$ for $J=1$.  
(a) Projectors of both observables taken as $\ket{1}\bra{1}$.  
(b) Projectors of both observables taken as $\ket{0}\bra{0}$.  
In both panels, curves for different system sizes are indicated in the legend.
}
\label{T_QSL}
\end{figure}

\subsection*{Numerical evaluation}

For completeness, we describe the numerical procedure used to evaluate
Eq.~\eqref{eq:QSL_time_nonpositivity}. For each choice of system parameters,
sites $(m,n)$, and projector outcomes $(\gamma,\delta)$, we first construct
the Hamiltonian $H$, its initial ground state
$\rho_0=\ket{\Psi_0}\bra{\Psi_0}$, and the embedded projectors
$\Pi_\gamma^m$ and $\Xi_\delta^n$. We then form
$\rho_\delta^n=\{\rho_0,\Xi_\delta^n\}/2$ and diagonalize it to obtain its
extremal eigenvalues $x_{\min}$ and $x_{\max}$. The initial normalized MH
entry is evaluated as
$\widetilde q_{\gamma\delta}^{mn}(0)
=\operatorname{Tr}[
\rho_\delta^n\widetilde{\Pi}_\gamma^m]$.

The SLD can be obtained by diagonalizing
$\widetilde{\Pi}_\gamma^m
=\mathcal{V}\operatorname{diag}(p_a)\mathcal{V}^\dagger$ and evaluating
$\dot{\widetilde{\Pi}}_\gamma^m
=i[H,\widetilde{\Pi}_\gamma^m]$. In this eigenbasis,
\begin{equation}
\left(
L_\gamma^m
\right)_{ab}
=
\begin{cases}
\dfrac{
2
\left(
\dot{\widetilde{\Pi}}_\gamma^m
\right)_{ab}
}{
p_a+p_b
},
&
p_a+p_b\neq0,
\\[3mm]
0,
&
p_a+p_b=0.
\end{cases}
\label{eq:QSL_SLD_matrix_elements}
\end{equation}
The operator is then transformed back to the original basis and its variance
is evaluated using Eq.~\eqref{eq:QSL_SLD_variance}.

For the normalized projector used here, the SLD admits the simpler expression
\begin{equation}
L_\gamma^m
=
2i
\left[
H,\Pi_\gamma^m
\right],
\label{eq:QSL_projector_SLD}
\end{equation}
and hence
\begin{equation}
\left(
\Delta L_\gamma^m
\right)^2
=
\frac{4}{r_\gamma}
\operatorname{Tr}
\left[
\Pi_\gamma^m
H
\left(
\mathbb{I}-\Pi_\gamma^m
\right)
H
\Pi_\gamma^m
\right].
\label{eq:QSL_projector_variance}
\end{equation}
Equation~\eqref{eq:QSL_projector_variance} is the expression used in the
numerical calculations because it is manifestly nonnegative and avoids an
explicit construction of the SLD.

Finally, we compute
$\tau(\rho_\delta^n,
\widetilde q_{\gamma\delta}^{mn}(0))$ and
$\tau(\rho_\delta^n,0)$ from
Eq.~\eqref{eq:QSL_interpolation_angle}, and substitute them into
Eq.~\eqref{eq:QSL_time_nonpositivity}. In the numerical implementation, the
argument of the arccosine is clipped to $[-1,1]$ only to remove floating-point
roundoff after verifying that the corresponding value lies within the
spectral interval.

For comparison with the FTN of the local $\sigma_z$ quasiprobability in the
integrable TFIM, the relevant entry is $q_{11}^{mm}(t)$. We therefore use
\begin{equation}
T_{\mathrm{QSL}}
=
T_{11;mm}^{\mathrm{re}}
\label{eq:TQSL_used}
\end{equation}
in Figures~\ref{tl_hx_auto} and \ref{FTN_l}. Figure~\ref{T_QSL} additionally presents
$T_{00;mm}^{\mathrm{re}}$ to demonstrate that a finite time to possible
nonpositivity does not imply that the associated MH entry actually becomes
negative.

For the projector pair $\ket{1}\!\bra{1}$ [Figure~\ref{T_QSL} (a)], the calculated
$T_{\mathrm{QSL}}$ follows the qualitative dependence of the observed FTN on
the transverse field while remaining below it. For the projector pair
$\ket{0}\!\bra{0}$ [Figure~\ref{T_QSL} (b)], the QSL can remain finite even though the corresponding MH
entry remains nonnegative in the regimes studied. This illustrates the
distinction between a geometrically allowed time to nonpositivity and the
actual dynamical generation of quasiprobability negativity.

\section{Position dependence of the FTN and QSL}
\label{FTN_position}
We investigate how the position of an on-site measurement affects the FTN and
the corresponding QSL in the integrable transverse-field Ising model with open
boundary conditions. Both projectors are placed at the same site,
$m=n=j$, and the site $j$ is varied across the chain.

\begin{figure}
    \centering
\includegraphics[width=0.32\linewidth,height=0.25\linewidth]{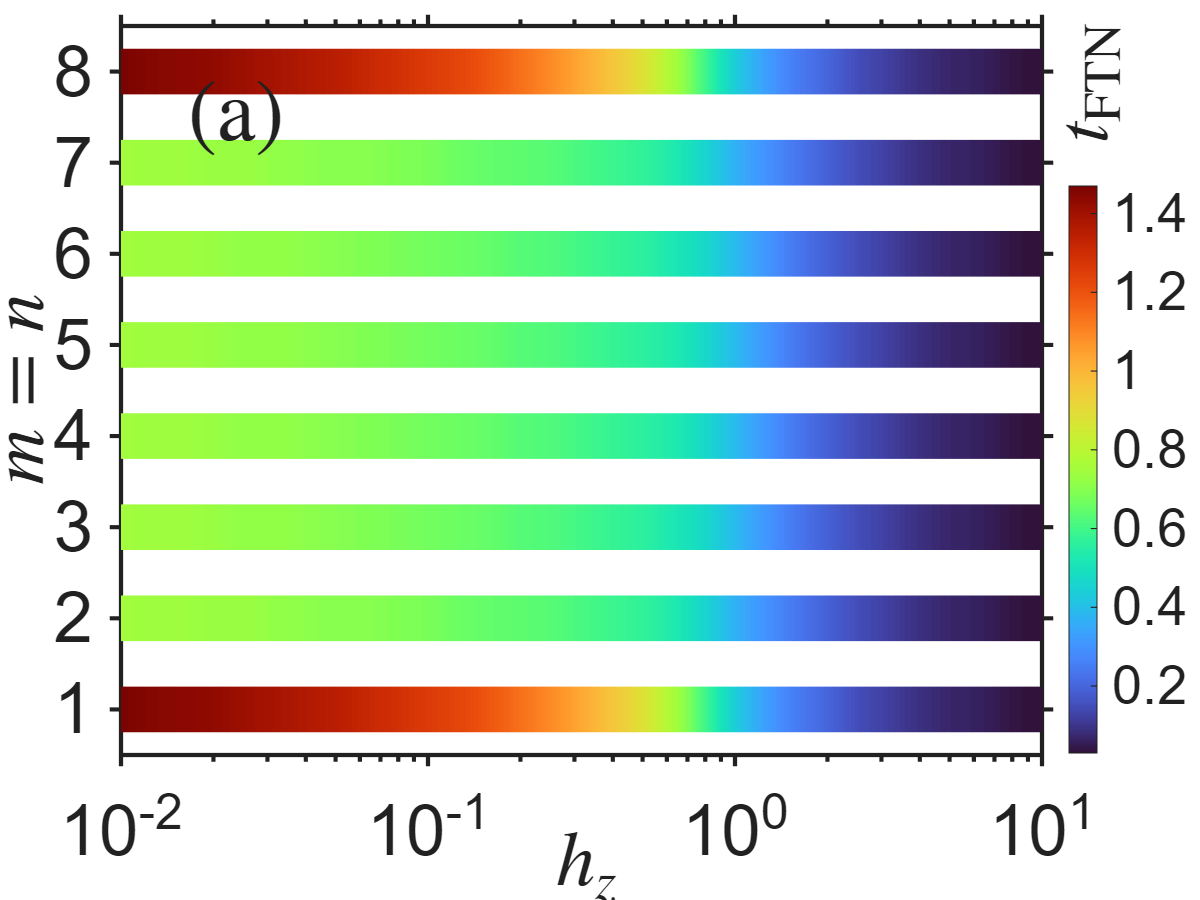}
\includegraphics[width=0.32\linewidth,height=0.25\linewidth]{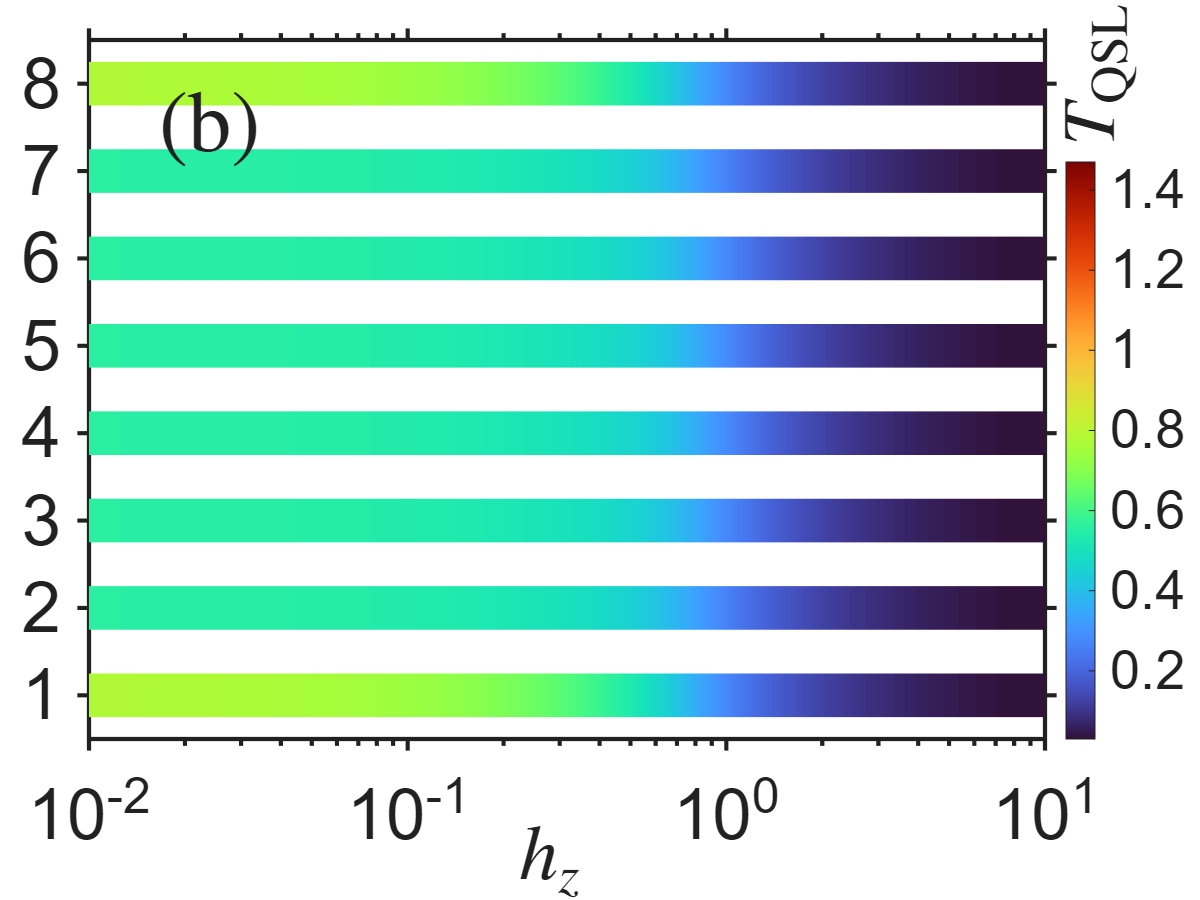}
\includegraphics[width=0.32\linewidth,height=0.25\linewidth]{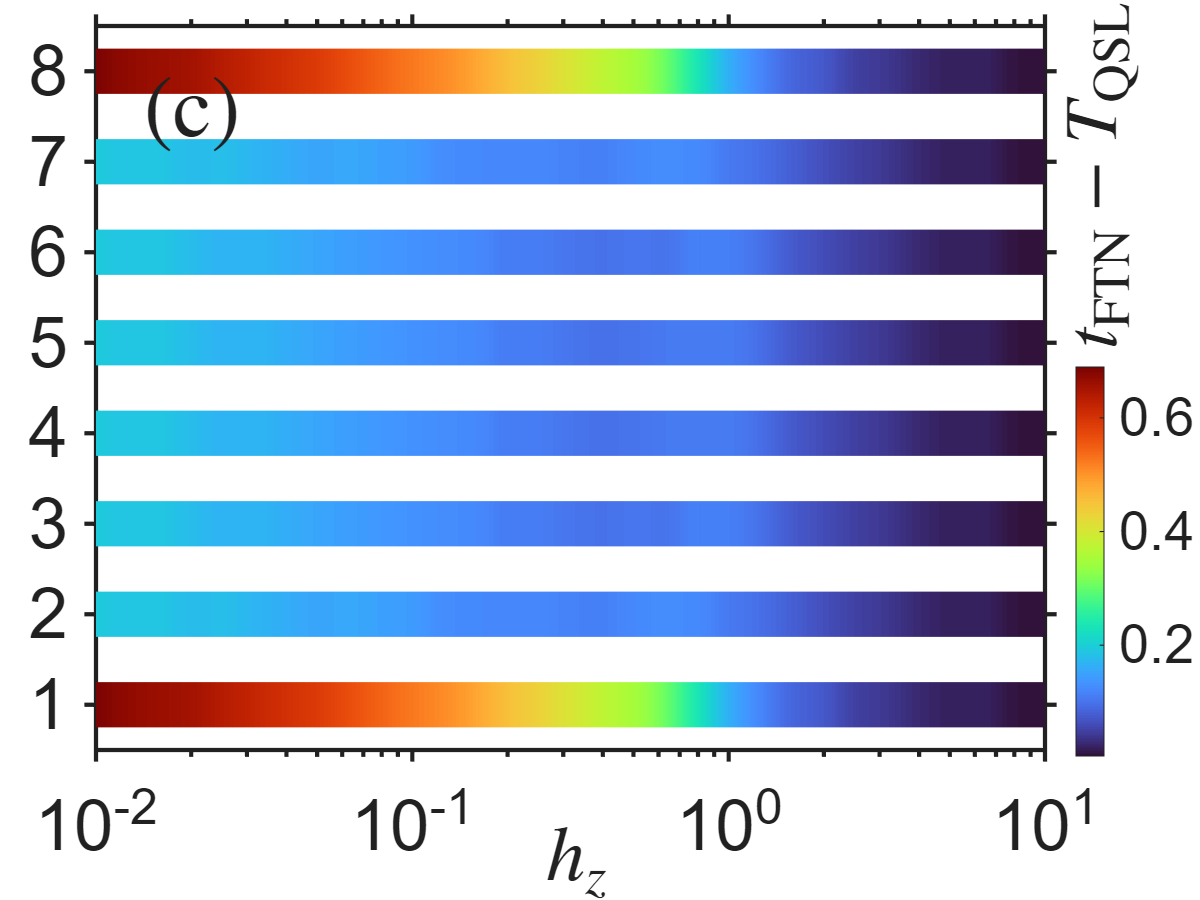}
\caption{
Local-site dependence of first-time negativity and the corresponding QSL time
in the integrable transverse-field Ising model \((h_x=0)\) with open boundary
conditions. The vertical axis labels the measured site, with \(m=n\), and the
horizontal axis is the transverse field \(h_z\) on a logarithmic scale
(\(J=1\)). Each horizontal row corresponds to one discrete lattice site; no
interpolation is used between different sites. 
(a) First-time negativity \(t_{\rm FTN}\). 
(b) Quantum-speed-limit time \(T_{\rm QSL}\), which gives a lower bound on the
time at which quasiprobability nonpositivity can occur. 
(c) Difference \(t_{\rm FTN}-T_{\rm QSL}\), showing how close the bound is to
the actual observed first-time negativity. 
All times are given in units of \(1/J\). The enhanced values near the edges
reflect the open-boundary geometry, while the bulk rows show the weaker
site-dependence away from the boundaries.
}
\label{FTN_l}
\end{figure}

Figure~\ref{FTN_l}(a) shows that the main spatial dependence is a
boundary-bulk contrast. In the interaction-dominated regime, $h_z<J$, the
FTN at the two boundary sites is larger than in the bulk, whereas the bulk
sites exhibit only a weak position dependence. In the field-dominated regime,
the difference between boundary and bulk becomes substantially smaller. The
two boundaries give the same result, as expected from the reflection symmetry
of the open chain.

The boundary-bulk contrast arises because a local projector at the edge and
one in the bulk experience different local environments and couple
differently to the excitations of the system. This can shift the first
zero-crossing of the MH quasiprobability and therefore modify the observed
FTN.

To compare the actual onset with the geometric speed bound, we evaluate
$T_{\mathrm{QSL}}$ for the projector pair $\ket{1}\!\bra{1}$ at each on-site
position, following the procedure described in App.~\ref{App:QSL}.
Figure~\ref{FTN_l}(b) shows a similar boundary-bulk contrast: for the
parameters considered, the boundary sites exhibit larger
$T_{\mathrm{QSL}}$ than the bulk sites, while the QSL remains below the
corresponding FTN.

The difference $t_{\mathrm{FTN}}-T_{\mathrm{QSL}}$, shown in
Fig.~\ref{FTN_l}(c), quantifies the separation between the earliest onset
permitted by the geometric bound and the actual dynamical first-time
negativity. These results show that, for open boundary conditions, the
location of the local measurement can affect both the observed onset of MH
negativity and its QSL benchmark, with the dominant distinction occurring
between boundary and bulk sites.

\section{ Asymptotic scaling of the FTN}
\label{app:asymptotic_scale}
An analytical expression for the quasiprobability for arbitrary system size
$N$ in the integrable transverse-field Ising model is derived in
App.~\ref{ana_calc}. This expression contains the full many-body spectrum
through the Bogoliubov modes and is used to compute the many-body FTN shown in
Fig.~\ref{tl_hx_auto}. The numerical data show that, over the system sizes
studied, the first onset time depends only weakly on $N$ and exhibits the same
asymptotic trends in both limits $h_z\ll J$ and $h_z\gg J$.

To obtain a transparent analytical explanation of these asymptotic trends, we therefore consider a minimal two-qubit model. This model is not intended to
describe thermodynamic-limit relaxation, dephasing, or long-time recurrences.
Rather, it isolates the local short-time mechanism that controls the first zero crossing of the Margenau-Hill quasiprobability and identifies the relevant energy scales governing $t_{\rm FTN}$. We emphasize that the use of this minimal model does not constitute a rigorous proof of the thermodynamic-limit scaling; deriving an exact analytical expression for
$t_{\rm FTN}$ as $N\to\infty$ remains an open theoretical problem.
%
The corresponding two-qubit Hamiltonian is given by:

\begin{eqnarray}
{H} &&= -J  {\sigma}_1^x \otimes {\sigma}_{2}^x - h_z \left( \hat{\sigma}_1^z \otimes \mathbb{ I} +\mathbb{ I} \otimes {\sigma}_2^z\right) \nonumber \\ 
&&= \begin{bmatrix}
-2h_z & 0 & 0 &-J \\
0 & 0 & -J & 0 \\
0 & -J & 0 &0 \\
-J & 0 & 0 & 2h_z 
\end{bmatrix}.
\label{Ham_toy}
\end{eqnarray}
Diagonalization of this Hamiltonian yields the eigenvalues $E$ and eigenvectors $V$, which are essential for constructing the time evolution operator and the ground state:
\begin{eqnarray}
E&&=\begin{pmatrix}
-J & 0 & 0 & 0 \\
0 & J & 0 & 0 \\
0 & 0 & -\sqrt{4 h_z^2 + J^2} & 0 \\
0 & 0 & 0 & \sqrt{4 h_z^2 + J^2}
\end{pmatrix}, \nonumber \\
V&& = 
\begin{pmatrix}
0 & 0 & -\frac{-\sqrt{4 h_z^2 + J^2} - 2 h_z}{J} & -\frac{\sqrt{4 h_z^2 + J^2} - 2 h_z}{J} \\
1 & -1 & 0 & 0 \\
1 & 1 & 0 & 0 \\
0 & 0 & 1 & 1
\end{pmatrix}. \nonumber 
\end{eqnarray}
The spectrum consists of four energy levels, \(\pm J\) and \(\pm \sqrt{4 h_z^2 + J^2}\), with the lowest energy level
\[
E_0 = -\sqrt{4 h_z^2 + J^2}
\]
and the corresponding eigenvector defining the normalized ground state \(\ket{\Psi_0}\):
\begin{eqnarray}
\ket{ \Psi_{0}} &&= 
\begin{bmatrix}
\frac{2 h_z + \sqrt{4 h_z^2 + J^2}}{ \sqrt{\left( \sqrt{4 h_z^2 + J^2} - 2 h_z \right)^2 + J^2}}\\ 0\\ 0\\ \frac{J}{\sqrt{\left( \sqrt{4 h_z^2 + J^2} + 2 h_z \right)^2 + J^2}}
\end{bmatrix}.
\end{eqnarray}
The time evolution is governed by the unitary operator \(U(t) = e^{-i H t}\). Its matrix representation in the computational basis is:

\begin{eqnarray}
&& U(t) =\nonumber \\
&&\left(
\begin{array}{cc}
\cos \left(t \sqrt{4 h_z^2 + J^2}\right) + \frac{2 i h_z \sin \left(t \sqrt{4 h_z^2 + J^2}\right)}{\sqrt{4 h_z^2 + J^2}} & 0 \\
0 & \cos (J t) \\
0 & i \sin (J t) \\
\frac{i J \sin \left(t \sqrt{4 h_z^2 + J^2}\right)}{\sqrt{4 h_z^2 + J^2}} & 0 
\end{array}
\right.
\nonumber \\
&&\left.
\begin{array}{cc}
0 & \frac{i J \sin \left(t \sqrt{4 h_z^2 + J^2}\right)}{\sqrt{4 h_z^2 + J^2}} \\
i \sin (J t) & 0 \\
\cos (J t) & 0 \\
0 & \cos \left(t \sqrt{4 h_z^2 + J^2}\right) - \frac{2 i h_z \sin \left(t \sqrt{4 h_z^2 + J^2}\right)}{\sqrt{4 h_z^2 + J^2}}
\end{array}
\right) \nonumber 
\end{eqnarray}
In the integrable TFIM, only the Pauli observables \(V = W = \sigma^z_n\) contribute to negativity. Focusing on the boundary positions, \(V = W = \sigma^z_1\), it is found that a single combination of projectors, \(\ket{1}\bra{1}\), yields negative quasiprobability. This indicates that only this quasiprobability contributes to the calculation of FTN. The corresponding projectors are
\begin{equation}
{\Xi}_1^1 = {\Pi}_1^1 =
\begin{bmatrix}
0 & 0 & 0 & 0\\
0 & 0 & 0 & 0\\
0 & 0 & 1 & 0\\
0 & 0 & 0 & 1\\
\end{bmatrix},
\end{equation}
with the quasiprobability defined as
\begin{equation}
q_{11}^{11}(t) = \bra{\Psi_0}{\Pi}_1^1(t) \, {\Xi}_1^1 \, \ket{\Psi_0}.
\end{equation}
After simplification, the real part is expressed as
\begin{equation}
\Re[q_{11}^{11}(t)] = \frac{4 h_z^2 - 2 h_z \sqrt{4 h_z^2 + J^2} + J^2 \cos^2\left( \sqrt{4 h_z^2 + J^2} \, t \right)}{8 h_z^2 + 2 J^2}.
\label{P22_neg} 
\end{equation}
This provides the condition for the FTN, \(\Re[q_{11}^{11}(t)] \le 0\), and allows analysis of its asymptotic behavior as the transverse field varies.  

\paragraph{Case I: \(h_z \ll J\):}  
Defining the small parameter \(\epsilon = h_z / J \ll 1\), expansion of the square root yields \(\sqrt{4 \epsilon^2 + 1} \approx 1 + 2 \epsilon^2\). Substituting into Eq.~(\ref{P22_neg}) and solving for the time gives
\begin{equation}
t_{\rm FTN} \equiv t \approx \frac{\pi}{2J} - \sqrt{\frac{2 h_z}{J^3}},
\label{hz_l_J}
\end{equation}
highlighting that the FTN is dominated by \(1/J\) with a subleading correction proportional to \(\sqrt{h_z / J^3}\).

\paragraph{Case II: \(h_z \gg J\):}  
For \(\epsilon = J / h_z \ll 1\), the condition \(\Re[q_{11}^{11}(t)] = 0\) leads to
\begin{equation}
t_{\rm FTN} \equiv t \approx \frac{\pi h_z}{J^2 + 8 h_z^2} \approx \frac{\pi}{8 h_z},
\label{hz_g_J}
\end{equation}
showing that in the strong-field regime, the FTN scales asymptotically as \(1/h_z\).  

\begin{figure}
    \centering
\includegraphics[width=0.60\linewidth,height=0.4\linewidth]{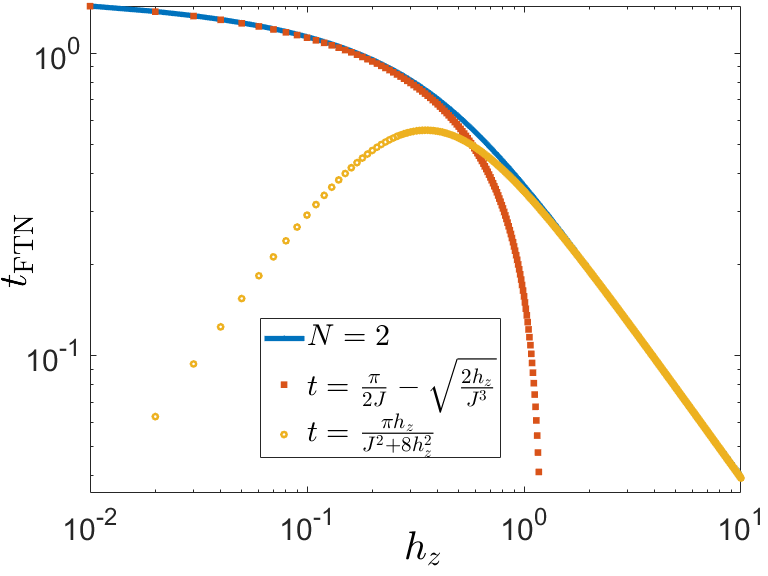}
\caption{
Exact and asymptotic behavior of the first-time negativity in the transverse-field Ising model for $N=2$ and $J=1$.  
The blue line shows the exact $t_{\rm FTN}$ as a function of the transverse field $h_z$.  
Orange squares indicate the weak-field approximation 
$t_{\rm FTN} = \frac{\pi}{2J} - \sqrt{2h_z/J^3}$ (valid for $h_z \ll J$),  
and yellow circles indicate the strong-field approximation 
$t_{\rm FTN} = \frac{\pi h_z}{J^2 + 8h_z^2}$ (valid for $h_z \gg J$).
}
\label{Num_ana}
\end{figure}

The asymptotic expressions in Eqs.~(\ref{hz_l_J}) and (\ref{hz_g_J}) reproduce the
weak- and strong-field scaling of $t_{\rm FTN}$ and agree with the exact
two-qubit numerics shown in Fig.~\ref{Num_ana}.

\medskip

In the strong-field regime ($J\ll h_z$), not only does the FTN scale as
$t_{\rm FTN}\sim 1/h_z$, but the \emph{magnitude} of the quasiprobability
oscillations is also parametrically suppressed. To see this explicitly, expand
the frequency $\Omega=\sqrt{4h_z^{2}+J^{2}}$ appearing in Eq.~(\ref{P22_neg}) as
\begin{equation}
\Omega
=2h_z\sqrt{1+\frac{J^2}{4h_z^2}}
=2h_z+\frac{J^{2}}{4h_z}
+\mathcal{O}\!\left(\frac{J^{4}}{h_z^{3}}\right).
\end{equation}
Substituting into the numerator of Eq.~(\ref{P22_neg}) gives,
$
4h_z^{2}-2h_z\Omega
= -\,\frac{J^{2}}{2}
+\mathcal{O}\!\left(\frac{J^{4}}{h_z^{2}}\right),
$
and therefore
\begin{equation}
\Re\!\left[q^{11}_{11}(t)\right]
\simeq \frac{J^{2}}{8h_z^{2}}
\left[\cos^{2}(2h_z t)-\frac{1}{2}\right],
\qquad (J\ll h_z).
\label{eq:qh_strongfield_amp}
\end{equation}
Equation~\eqref{eq:qh_strongfield_amp} makes two points transparent:
(i) the oscillation amplitude scales as $J^{2}/h_z^{2}$ and thus becomes
progressively smaller at larger $h_z$, making the resulting negativity
increasingly difficult to resolve in finite-precision numerics or in the
presence of experimental noise; (ii) for any finite $J\neq 0$ the bracketed term
attains negative values (since $\cos^2(2h_zt)<1/2$ over finite time intervals),
so the quasiprobability still becomes negative at some times even though the
negativity is parametrically weak.

\section{ Quasiprobability formula for a spin chain of arbitrary length $N$}
\label{ana_calc}
The integrable TFIM can be diagonalized, which allows us to obtain the exact formula of quasiprobability. To do this, we first use the relations $\sigma^{x/z}_n = 2 S^{x/z}_n$ and $S^{\pm}_n = S^x_n \pm i S^y_n$, so that $S^x_n = \tfrac{1}{2}(S^+_n + S^-_n)$. Substituting $S^x_n$ and $h_x = 0$ into Eq.~\eqref{eq:H}, the spin Hamiltonian reduces to
\begin{eqnarray}
H &&= -J \sum_{n=1}^{N-1} \left( S^+_n S^+_{n+1} + S^+_n S^-_{n+1} + S^-_n S^+_{n+1} + S^-_n S^-_{n+1} \right) \nonumber \\
&&\quad - 2 h_z \sum_{n=1}^{N} S^z_n.
\end{eqnarray}
The spin operators can be expressed in terms of fermionic operators through the Jordan–Wigner transformation:
\begin{eqnarray}
S_n^- &=& \exp\left( i\pi\sum_{l=1}^{n-1} c_l^\dagger c_l \right) c_n, \nonumber \\
S_n^+ &=& c_n^\dagger \exp\left( -i\pi\sum_{l=1}^{n-1} c_l^\dagger c_l \right), \nonumber \\
S_n^z &=& c_n^\dagger c_n - \frac{1}{2}.
\end{eqnarray}
For convenience, we introduce the non-local string operator
\begin{eqnarray}
Q_n = \exp\left( i\pi \sum_{l=1}^{n-1} c_l^\dagger c_l \right) = \prod_{l=1}^{n-1} \left( 1 - 2 c_l^\dagger c_l \right),
\end{eqnarray}
which satisfies the properties $Q_n^\dagger = Q_n$, $Q_n^2 = 1$, and $Q_n Q_{n+1} = 1 - 2 c_n^\dagger c_n$. Using this construction, the nearest-neighbor spin interactions can be expressed in fermionic form as
\begin{eqnarray}
&& S_n^+ S_{n+1}^+ = c_n^\dagger c_{n+1}^\dagger \nonumber \\
&&S_n^+ S_{n+1}^- = c_n^\dagger c_{n+1}
\nonumber \\
&&S_n^- S_{n+1}^+ = -c_n c_{n+1}^\dagger
\nonumber \\
&&S_n^- S_{n+1}^- = -c_n  c_{n+1}. \nonumber
\end{eqnarray}
Thus, the spin Hamiltonian is fully mapped into a fermionic representation, yielding the compact quadratic form
\begin{eqnarray}
H &=& -J \sum_{n=1}^{N-1} \left( c_n^\dagger c_{n+1} + c_{n+1}^\dagger c_n + c_n^\dagger c_{n+1}^\dagger + c_{n+1} c_n \right) \nonumber \\
&& - h_z \sum_{n=1}^{N} \left( 2 c_n^\dagger c_n - 1 \right),
\label{fermionic_Hamiltonian}
\end{eqnarray}
Since the Hamiltonian  is quadratic in the fermionic operators, it can be diagonalized using the Bogoliubov–Valatin (BV) transformation \cite{lieb1961two,d2021topological,xiao2009theory,he2017boundary}:
\begin{equation}
    c_n = \sum_k \left( A_{nk} b_k + B_{nk} b_k^\dagger \right), \label{BV_trans}
\end{equation}
where $b_k$ and $b_k^\dagger$ are new fermionic annihilation and creation operators. The matrices $A$ and $B$ (discussed later how to construct them), obtained via the Lieb–Schultz–Mattis (LSM) procedure, bring the Hamiltonian into the diagonal form
\begin{equation}
  H = \sum_k \omega_k b_k^\dagger b_k + \mathrm{const}, \label{Dig_ham}
\end{equation}
with a real, nonnegative excitation spectrum given by
\begin{equation}
   \omega_k^2 = 4J^2 + 8Jh_z \cos(k) + 4h_z^2, \label{omega_express}
\end{equation}
The allowed values of $k$ are determined by the boundary condition
\begin{equation}
 J \sin(k N) + h_z \sin[k(N + 1)] = 0.
 \label{Quantization conditions}
\end{equation}
Thus, the diagonalization requires both the eigenvalues $\omega_k$, Eq.~\eqref{omega_express}, and the quantization condition for $k$, Eq.~\eqref{Quantization conditions}. In what follows, we outline the procedure to obtain these results.

{\it Lieb–Schultz–Mattis Method:} We are employing the LSM method to diagonalize the quadratic fermionic Hamiltonian, Eq.~(\ref{fermionic_Hamiltonian}), which is defined in terms of the real matrices $Q_{mn}$ and $P_{mn}$:
\begin{equation}
H = \sum_{m,n=1}^{N} \left[ Q_{mn} c_m^\dagger c_n + \frac{1}{2} P_{mn} (c_m^\dagger c_n^\dagger + c_n c_m) \right] + h_z N, 
\end{equation}
where the matrices $Q_{mn}$ and $P_{mn}$ are given explicitly by
\begin{align}
Q_{mn} &=
\begin{bmatrix}
-2h_z & -J & 0 & \cdots & 0 \\
-J & -2h_z & -J & \cdots & 0 \\
0 & -J & -2h_z & \cdots & 0 \\
\vdots & \vdots & \vdots & \ddots & \vdots \\
0 & 0 & 0 & \cdots & -2h_z
\end{bmatrix}, \\
P_{mn} &=
\begin{bmatrix}
0 & -J & 0 & \cdots & 0 \\
J & 0 & -J & \cdots & 0 \\
0 & J & 0 & \cdots & 0 \\
\vdots & \vdots & \vdots & \ddots & \vdots \\
0 & 0 & 0 & \cdots & 0
\end{bmatrix}.
\end{align}
The inverse BV transformation, with $X = A^T$ and $Y = B^T$, reads
\begin{equation}
   b_k = \sum_{n=1}^N \left( X_{kn} c_n + Y_{kn} c_n^\dagger \right), \label{inv_BV_transform}
\end{equation}
Expressing this in terms of the vectors ${\bf X}_k^T = (X_{k1}, \ldots, X_{kN})$ and ${\bf Y}_k^T = (Y_{k1}, \ldots, Y_{kN})$ leads to the consistency relations
\begin{subequations}
\begin{align}
Q {\bf X}_k + P {\bf Y}_k &= \omega_k {\bf X}_k, \label{eq:11a} \\
- Q {\bf X}_k - P {\bf Y}_k &= \omega_k {\bf Y}_k. \label{eq:11b}
\end{align}
\end{subequations}
Introducing $\bm{\phi}_k = {\bf X}_k + {\bf Y}_k$ and $\bm{\psi}_k = {\bf X}_k - {\bf Y}_k$, these equations can be recast as
\begin{subequations}
\begin{align}
(Q + P) \bm{\phi}_k &= \omega_k \bm{\psi}_k, \\
(Q - P) \bm{\psi}_k &= \omega_k \bm{\phi}_k.
\end{align}
\end{subequations}
From this, a standard eigenvalue problem emerges:
\begin{subequations}
\begin{align}
V \bm{\phi}_k &= \omega_k^2 \bm{\phi}_k, \\
W \bm{\psi}_k &= \omega_k^2 \bm{\psi}_k,
\end{align}
\end{subequations}
with $V = (Q - P)(Q + P)$ and $ W = (Q + P)(Q - P)$, these can be expressed in matrix form as follows:
\begin{eqnarray}
    &&V =
\begin{pmatrix}
4h_z^2 & 4Jh_z & 0 & \cdots & 0 \\
4Jh_z & 4J^2 + 4h_z^2 & 4Jh_z & \cdots & 0 \\
0 & 4Jh_z & 4J^2 + 4h_z^2 & \cdots & 0 \\
\vdots & \vdots & \vdots & \ddots &4Jh_z \\
0 & 0 & 0 & 4Jh_z & 4J^2 + 4h_z^2
\end{pmatrix},\nonumber \\
&&
W =
\begin{pmatrix}
4J^2 + 4h_z^2 & 4Jh_z & 0 & \cdots & 0 \\
4Jh_z & 4J^2 + 4h_z^2 & 4Jh_z & \cdots & 0 \\
0 & 4Jh_z & 4J^2 + 4h_z^2 & \cdots & 0 \\
\vdots & \vdots & \vdots & \ddots & 4Jh_z \\
0 & 0 & 0 & 4Jh_z & 4h_z^2
\end{pmatrix} \nonumber.
\end{eqnarray}
From the structure of these matrices, it follows that $\bm{\phi}_k$ is related to $\bm{\psi}_k$ through an index inversion $n \to N+1-n$. Hence, it is sufficient to solve for $\bm{\psi}_k$ by determining the eigenvalues and eigenvectors of $W$. The corresponding eigenvalue equation leads to the following system:
\begin{subequations}
\begin{align}
&4J h_z (\bm{\psi}_k)_{n-1} + (4J^2 + 4h_z^2) (\bm{\psi}_k)_n + 4J h_z (\bm{\psi}_k)_{n+1}
= \nonumber  \\ & \quad \quad \omega_k^2 (\bm{\psi}_k)_n, \quad \quad \quad \quad 2 \le n \le N-1, \label{boundary_equation:a} \\
&(4J^2 + 4h_z^2) (\bm{\psi}_k)_1 + 4J h_z (\bm{\psi}_k)_2
= \omega_k^2 (\bm{\psi}_k)_1, \label{boundary_equation:b} \\
&4J h_z (\bm{\psi}_k)_{N-1} + 4 h_z^2 (\bm{\psi}_k)_N
= \omega_k^2 (\bm{\psi}_k)_N. \label{boundary_equation:c}
\end{align}
\end{subequations}
Equation~\eqref{boundary_equation:a} governs the bulk dynamics, while Eqs.~\eqref{boundary_equation:b} and \eqref{boundary_equation:c} enforce the left and right boundary conditions, respectively. To solve the bulk equation, we assume a translationally invariant ansatz
\begin{align}
        (\bm{\psi}_k)_n = -i C_k \frac{1}{2} \left( e^{i k n} + \alpha_k e^{-i k n} \right),
 \end{align}
where $k$ is the quantum number (possibly complex), and $C_k$, $\alpha_k$ are parameters. Using the boundary conditions, one finds $\alpha_k = -1$, leading to
\begin{subequations}
    \begin{align}
\bm\psi_{nk} &= C_k \sin(k n), \label{psi_nk} \\
\bm\phi_{nk} &= D_k \sin[k(N + 1 - n)], \label{phi_nk}
\end{align}
\end{subequations}
where $C_k$ and $D_k$ are normalization constants related through
\begin{subequations}
\begin{align}
C_k^2 &= D_k^2, \\
\frac{2}{D_k^2} &= N - \frac{\sin(kN)}{\sin k} \cos[k(N+1)], \\
\frac{D_k}{C_k} &= - \frac{2 h_z \sin k}{\omega_k \sin(kN)}.
\end{align}
\label{Ck_Dk_relation}
\end{subequations}
These conditions imply that $C_k = \pm D_k$, with the relative sign depending on $k$, $h_z$, and $N$.  
Substituting the ansatz for $(\bm\psi_k)_n$ in Eq.~(\ref{boundary_equation:a}) yields the eigenvalues $\omega_k^2$ as in Eq.~\eqref{omega_express}, while the substituting the ansatz for $(\bm\psi_k)_n$ and value of  $\omega_k^2$ in right boundary condition, Eq.~\eqref{boundary_equation:c}, leads to the quantization condition defined in Eq.~\eqref{Quantization conditions}.

The time evolution operator is
\[
U = e^{-i H t} = e^{-i \sum_k \omega_k b_k^\dagger b_k t}.
\]
In the integrable TFIM, only the Pauli operators $V = W = \sigma_n^z$ contribute to negative values of the quasiprobability. Among all possible combinations of projectors for these operators, only the combination where both projectors are $\ket{1}\bra{1}$ yields a negative quasiprobability. Consequently, this is the only quasiprobability that contributes to the calculation of the FTN. The projectors, expressed in terms of fermionic operators, are given by
\begin{eqnarray}
\Pi^1_n \equiv \Xi^1_n= \frac{1}{2} (\mathbb{I} - \sigma_n^z) =\mathbb{I} - c_n^\dagger c_n=c_nc_n^\dagger.
\end{eqnarray}
Using these projectors, the quasiprobability can be expressed as
\begin{eqnarray}
q_{11}^{nn}(t)=&& \bra{\Psi_0}\Pi^n_1(t)\Xi^n_1 \ket{\Psi_0}=\bra{\Psi_0} c_n(t)  c_n^\dagger(t)c_n  c_n^\dagger\ket{\Psi_0}=  \nonumber \\
&&\bra{\Psi_0} \Big[\sum_{k,k'} \Big(
A_{nk} A_{nk'}^* \, b_{k}(t) b^\dagger_{k'}(t) 
+ A_{nk} B_{nk'}^* \, b_{k}(t) b_k'(t) \nonumber \\
&& \quad + B_{nk}^* A_{nk'}^* \, b_{k}^{\dagger}(t) b_{k'}^{\dagger} (t)
+ B_{nk} B_{nk'}^* \, b_{k}^{\dagger}(t) b_k'(t)
\Big)\Big] \nonumber \\
&&\quad \times \Big[ \sum_{p,p'} \Big(
A_{np} A_{np'}^* \, b_{p} b^\dagger_{p'}
+ A_{np} B_{np'}^* \, b_{p} b_p'
\Big) \nonumber \\
&& \quad + B_{np}^* A_{np'}^* \, b_{p}^{\dagger} b_{p'}^{\dagger}
+ B_{np} B_{np'}^* \, b_{p}^{\dagger} b_p'\Big]\ket{\Psi_0} 
\end{eqnarray}
The time-evolved fermionic creation and annihilation operators are obtained by:
\begin{align}
b_k^\dagger(t) = e^{i \omega_k t} b_k^\dagger,\quad {\rm and}\quad b_k(t) = e^{-i \omega_k t} b_k.
\end{align}
Taking the expectation value with respect to the fermionic ground state $|\Psi_0\rangle = \prod_{k, E_k \le E_F} b_k^\dagger |0\rangle$, and taking the real part, we have
\begin{align}
&\Re\left[ \langle \Psi_0 | \Pi^n_1(t) \Xi^n_1(0) | \Psi_0 \rangle \right] = \sum_{k,p} |A_{nk}|^2 |A_{np}|^2\nonumber \\ & 
+ \sum_{k,p} \left( |A_{np}|^2 |B_{nk}|^2 - A_{np}^* B_{nk} B_{np}^* A_{nk} \right) \cos\left[ (\omega_p + \omega_k) t \right].
\end{align}
This equation is useful for the calculation of the quasiprobability for a general $N$.

To determine the time at which the quasiprobability first becomes negative,  
we set the real part of the expression equal to zero. This yields the condition  \[
\sum_{k,p} C_{k,p} \cos\left[ (\omega_p + \omega_k) t \right] 
= - \sum_{k,p} |A_{nk}|^2 |A_{np}|^2.
\]
where $
C_{k,p} = |A_{np}|^2 |B_{nk}|^2 - A_{np}^* B_{nk} B_{np}^* A_{nk}$. Since the time variable $t$ appears inside the summations through oscillatory terms of the form $(\omega_p + \omega_k)$, the resulting expression does not admit a closed-form solution for its time dependence. As a result, it is challenging to extract the precise behavior of the negativity onset as the external transverse field $h_z$ is varied.


\section{ First-Time Negativity in the Antiferromagnetic Regime}
\label{FTN_J_s}
In the \textit{integrable} transverse-field Ising chain ($h_x=0$), the sign of the
exchange coupling $J$ can be gauged away on a bipartite 1D lattice.  For an
open chain with nearest-neighbour $\sigma^x\sigma^x$ interactions, define the
unitary
\begin{equation}
U \;=\; \prod_{j\in\mathrm{even}} \sigma_j^{z}.
\end{equation}
Using $\sigma^z\sigma^x\sigma^z=-\sigma^x$ and $\sigma^z\sigma^z\sigma^z=\sigma^z$,
one finds that $U$ flips $\sigma_j^x\mapsto -\sigma_j^x$ on even sites while
leaving all $\sigma_j^z$ invariant.  Since each bond $(j,j{+}1)$ connects an
odd and an even site, every exchange term acquires a single minus sign,
\begin{equation}
U\big(\sigma_j^x\sigma_{j+1}^x\big)U^\dagger \;=\; -\,\sigma_j^x\sigma_{j+1}^x,
\qquad
U\big(\sigma_j^z\big)U^\dagger \;=\; \sigma_j^z,
\end{equation}
and therefore
\begin{equation}
U\,H(J,h_z,h_x{=}0)\,U^\dagger \;=\; H(-J,h_z,h_x{=}0).
\end{equation}
Thus $H(J,h_z,0)$ and $H(-J,h_z,0)$ are unitarily equivalent and have identical
spectra.  Moreover, for local $\sigma_z$ projectors (which commute with $U$),
the associated Heisenberg-evolved projectors and hence the KDQ are mapped into each other under the same transformation.
Consequently, the first-time negativity $t_{\mathrm{FTN}}$ is unchanged under
$J\to -J$ in the integrable model.  This equivalence does \emph{not} extend to
the nonintegrable case $h_x\neq 0$, because $U\sum_j\sigma_j^x U^\dagger =
\sum_j(-1)^j\sigma_j^x$ turns a uniform longitudinal field into a staggered one.

 \begin{figure}
    \centering
\includegraphics[width=0.49\linewidth,height=0.40\linewidth]{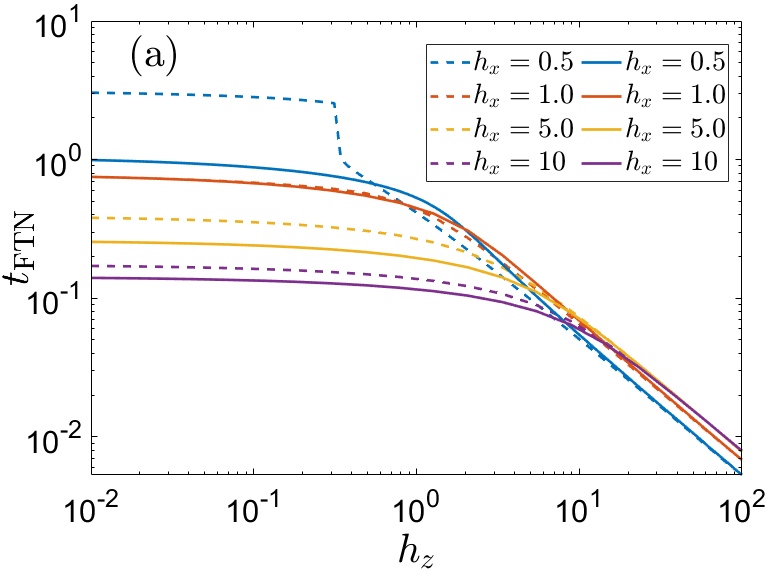}
\includegraphics[width=0.49\linewidth,height=0.40\linewidth]{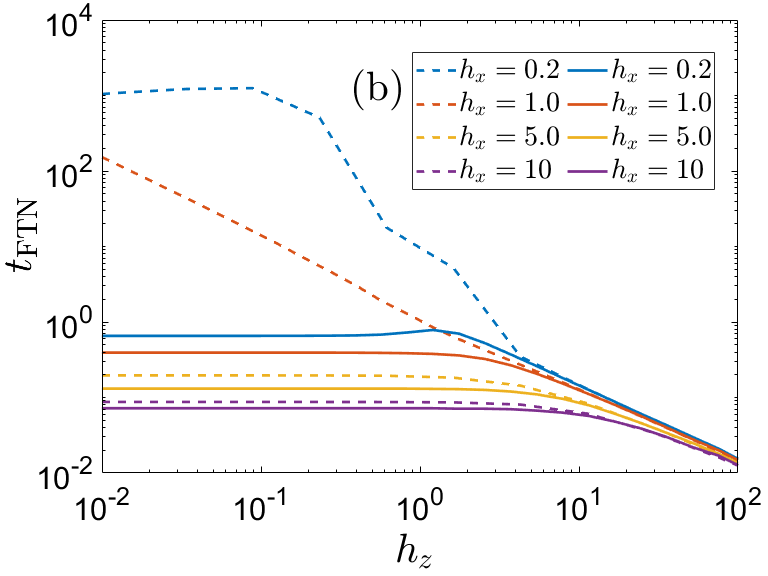}
 \caption{
First-time negativity $t_{\rm FTN}$ (in units of $1/J$) as a function of the transverse field $h_z$ for several fixed values of $h_x$ (see legends). Panels (a) correspond to the local observable $V=W=\sigma_z^1$, while panels (b) correspond to $V=W=\sigma_x^1$. Dashed lines indicate $J=-1$ (antiferromagnetic), whereas solid lines indicate $J=1$ (ferromagnetic). In all cases, the system size is $N=8$.
}
 \label{FTN_J}
\end{figure}

In the \emph{nonintegrable} case ($h_x\neq 0$), we compute $t_{\rm FTN}$ for the
local observables $\sigma_1^z$ and $\sigma_1^x$ and compare $J=1$ to $J=-1$.
For $\sigma_1^z$, the dependence on the sign of $J$ becomes weak already for
moderate longitudinal fields ($h_x\sim 1$) and is negligible for larger $h_x$.
For $\sigma_1^x$, by contrast, a pronounced sensitivity to the sign of $J$
persists in the low-field regime $h_z\lesssim |J|$ up to $h_x$ of order unity,
and only diminishes for stronger $h_x$ (see Fig.~\ref{FTN_J}).
In all cases the curves merge again at larger $h_z$, and in the strong-field
limit $h_z\gg |J|$ the FTN becomes essentially independent of the sign of $J$,
since the dynamics are dominated by the transverse field and the interaction
enters only as a subleading correction.

\section{Correlator thresholds for spatially separated measurements}
\label{app:spatial_thresholds}

For local $\sigma_z$ projectors,
\begin{equation}
\Pi_\gamma^m(t)
=
\frac{1}{2}
\left[
\mathbb{I}
+\lambda_\gamma\sigma_z^m(t)
\right],
\qquad
\Xi_\delta^n
=
\frac{1}{2}
\left[
\mathbb{I}
+\lambda_\delta\sigma_z^n
\right],
\end{equation}
with $\lambda_0=1$ and $\lambda_1=-1$. Substitution into
$q_{\gamma\delta}^{mn}(t)
=\operatorname{Re}\operatorname{Tr}
[\Pi_\gamma^m(t)\Xi_\delta^n\rho_0]$ gives
\begin{equation}
q_{\gamma\delta}^{mn}(t)
=
\frac{1}{4}
\left[
1
+\lambda_\gamma\langle\sigma_z^m(t)\rangle
+\lambda_\delta\langle\sigma_z^n\rangle
+\lambda_\gamma\lambda_\delta
\operatorname{Re}
C_{\sigma_z^m\sigma_z^n}(t)
\right].
\end{equation}
Since the initial ground state satisfies $[H,\rho_0]=0$,
$\langle\sigma_z^m(t)\rangle=\langle\sigma_z^m\rangle$, which yields
Eq.~\eqref{eq:spatial_MH}.

For the two equivalent boundaries,
$\langle\sigma_z^1\rangle=\langle\sigma_z^N\rangle$, and the four MH entries
reduce to
\begin{equation}
q_{11}^{1N}(t)
=
\frac{1}{4}
\left[
1-2\langle\sigma_z^1\rangle
+\operatorname{Re}
C_{\sigma_z^1\sigma_z^N}(t)
\right],
\end{equation}
\begin{equation}
q_{01}^{1N}(t)
=
q_{10}^{1N}(t)
=
\frac{1}{4}
\left[
1-\operatorname{Re}
C_{\sigma_z^1\sigma_z^N}(t)
\right],
\end{equation}
and
\begin{equation}
q_{00}^{1N}(t)
=
\frac{1}{4}
\left[
1+2\langle\sigma_z^1\rangle
+\operatorname{Re}
C_{\sigma_z^1\sigma_z^N}(t)
\right].
\end{equation}
For $h_z>0$, $\langle\sigma_z^1\rangle\geq0$, while
$-1\leq\operatorname{Re}C_{\sigma_z^1\sigma_z^N}(t)\leq1$. It follows that
$q_{01}^{1N}$, $q_{10}^{1N}$, and $q_{00}^{1N}$ remain nonnegative, whereas
$q_{11}^{1N}$ becomes negative when
\begin{equation}
\operatorname{Re}
C_{\sigma_z^1\sigma_z^N}(t)
<
2\langle\sigma_z^1\rangle-1.
\end{equation}
At $h_z=0$, the local transverse polarization vanishes and negativity would
require the correlator to be smaller than $-1$. This explains the absence of
opposite-boundary negativity at the zero-field endpoint.

In the weak-field ordered regime, a boundary $\sigma_z$ operator flips a
single boundary spin in the $x$-ordered basis and therefore has a leading
one-domain-wall contribution. The two opposite-boundary operators couple to
the same excitation sector. A bulk $\sigma_z$ operator instead creates two
domain walls at leading order, suppressing the corresponding boundary--bulk
correlator in this regime. The opposite-boundary geometry is therefore
distinguished both by its negativity threshold and by the excitation sector
contributing to its two-time correlator.

The maximal TFIM quasiparticle group velocity is
$v_{\max}=2\min(J,|h_z|)$~\cite{pfeuty1970one}. It gives the characteristic
propagation scales $(N-1)/(2|h_z|)$ for $|h_z|<J$ and
$(N-1)/(2J)$ for $|h_z|>J$.

Finally, the results are symmetric under reversal of the transverse field.
For $U_x=\prod_j\sigma_x^j$,
$U_xH(h_z)U_x^\dagger=H(-h_z)$ and
$U_x\sigma_z^jU_x^\dagger=-\sigma_z^j$. Consequently,
\begin{equation}
q_{\gamma\delta}^{mn}(t;-h_z)
=
q_{\bar{\gamma}\bar{\delta}}^{mn}(t;h_z),
\qquad
\bar{0}=1,\quad \bar{1}=0,
\end{equation}
and therefore
\begin{equation}
t_{\mathrm{FTN}}^{mn}(-h_z)
=
t_{\mathrm{FTN}}^{mn}(h_z).
\end{equation}
Field reversal exchanges the $q_{11}$ and $q_{00}$ entries but leaves the
first-time negativity unchanged.




\end{document}